\documentclass[aps,showpacs,showkeys,superscriptaddress]{revtex4}

\usepackage{graphicx}
\usepackage{makeidx}
\usepackage[]{hyperref}
\usepackage{amsmath}

\usepackage{amssymb,epsfig,graphics}
\usepackage{amscd}
\usepackage{verbatim}

\usepackage{amsmath}
\usepackage{amsthm}
\usepackage{amsfonts}
\usepackage{dsfont}

\begin{document}
\title{Electronic Properties of Carbon Nanostructures}

\author{Jan Smotlacha}\email{smota@centrum.cz}
\affiliation{Bogoliubov Laboratory of Theoretical Physics, Joint
Institute for Nuclear Research, 141980 Dubna, Moscow region, Russia}
\affiliation{Faculty of Nuclear Sciences and Physical Engineering, Czech Technical University, Brehova 7, 110 00 Prague,
Czech Republic}

\author{Richard Pincak}\email{pincak@saske.sk}
\affiliation{Institute of Experimental Physics, Slovak Academy of Sciences,
Watsonova 47,043 53 Kosice, Slovak Republic}
\affiliation{Bogoliubov Laboratory of Theoretical Physics, Joint
Institute for Nuclear Research, 141980 Dubna, Moscow region, Russia}

\date{\today}

\begin{abstract}
The carbon nanostructures are perspective materials for the future applications. This has two reasons: first, the hexagonal atomic structure which enables a high molecular variability by placing different kinds of the defects and second, good electronic properties which can be modified for the purpose of the concrete applications with the help of the defects and of the chemical ingredients. A lot of kinds of the nanostructures was investigated. Here, the properties of less common forms will be examined - the graphitic nanocone and graphitic wormhole. 
\end{abstract}


\maketitle

\section{Introduction}\

\begin{figure}[htbp]
\includegraphics[width=40mm]{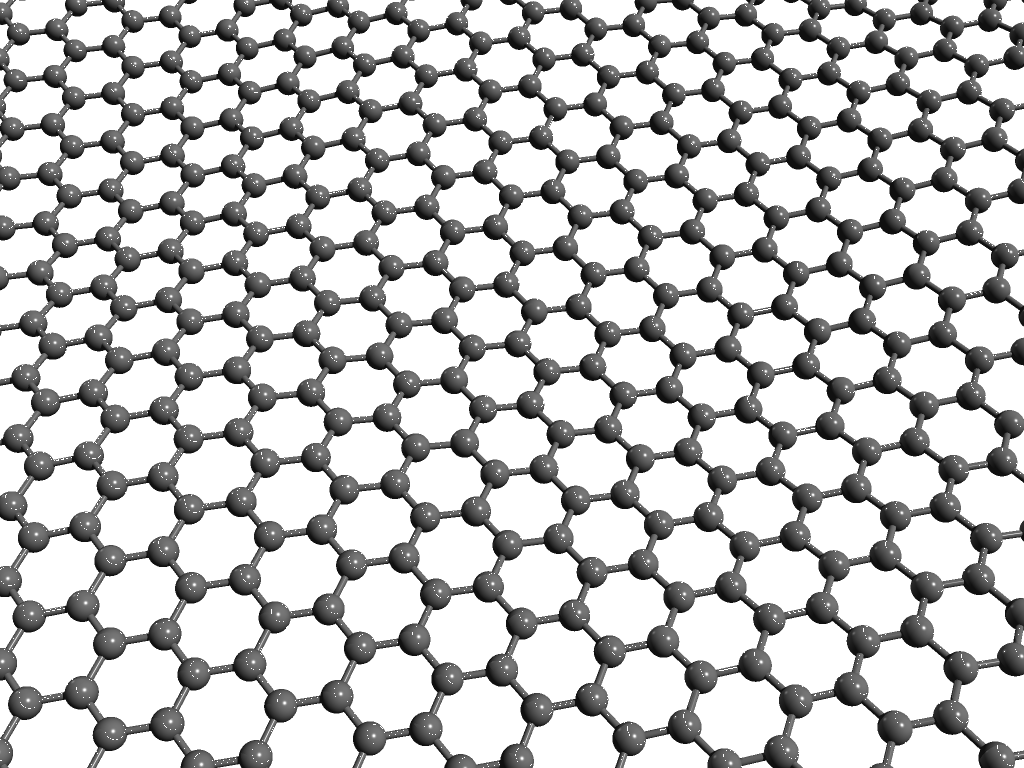}
\caption{Hexagonal carbon plain lattice.}\label{FG0}
\end{figure}

The carbon nanostructures are the materials whose molecular structure is derived from graphene - the hexagonal carbon plain lattice (Figure \ref{FG0}). Because of their electronic structure, they are the promising materials for the construction of nanoscale devices (quantum wires, nonlinear electronic
elements, transistors, molecular memory devices or electron field emitters) and the inventions in the material science.

\begin{figure}[htbp]
\includegraphics[width=120mm]{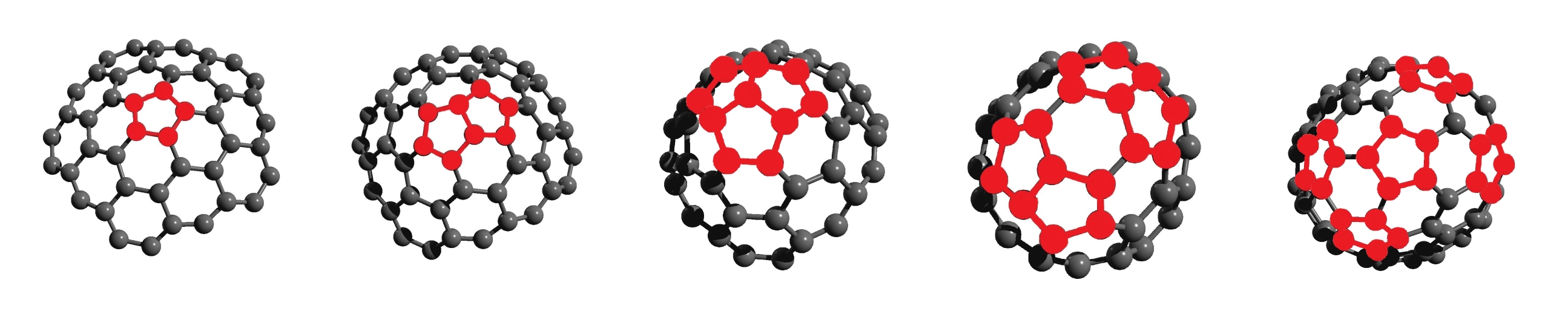}
\caption{Conical nanostructures with different numbers of pentagonal defects in the tip.}\label{PentCones}
\end{figure}

The planar geometry of the molecular surface is disrupted by the disclinations in the molecule structure which are most often presented by the pentagons and the heptagons in the hexagonal lattice. This change of the geometry is manifested by the positive or the negative curvature, respectively which can be enlarged by the supply of higher number of the defects. In this way, by the supply of 1 to 5 pentagonal defects, we get conical structures with different values of the vortex angle (Figure \ref{PentCones}).

\begin{figure}[htbp]
\includegraphics[width=35mm]{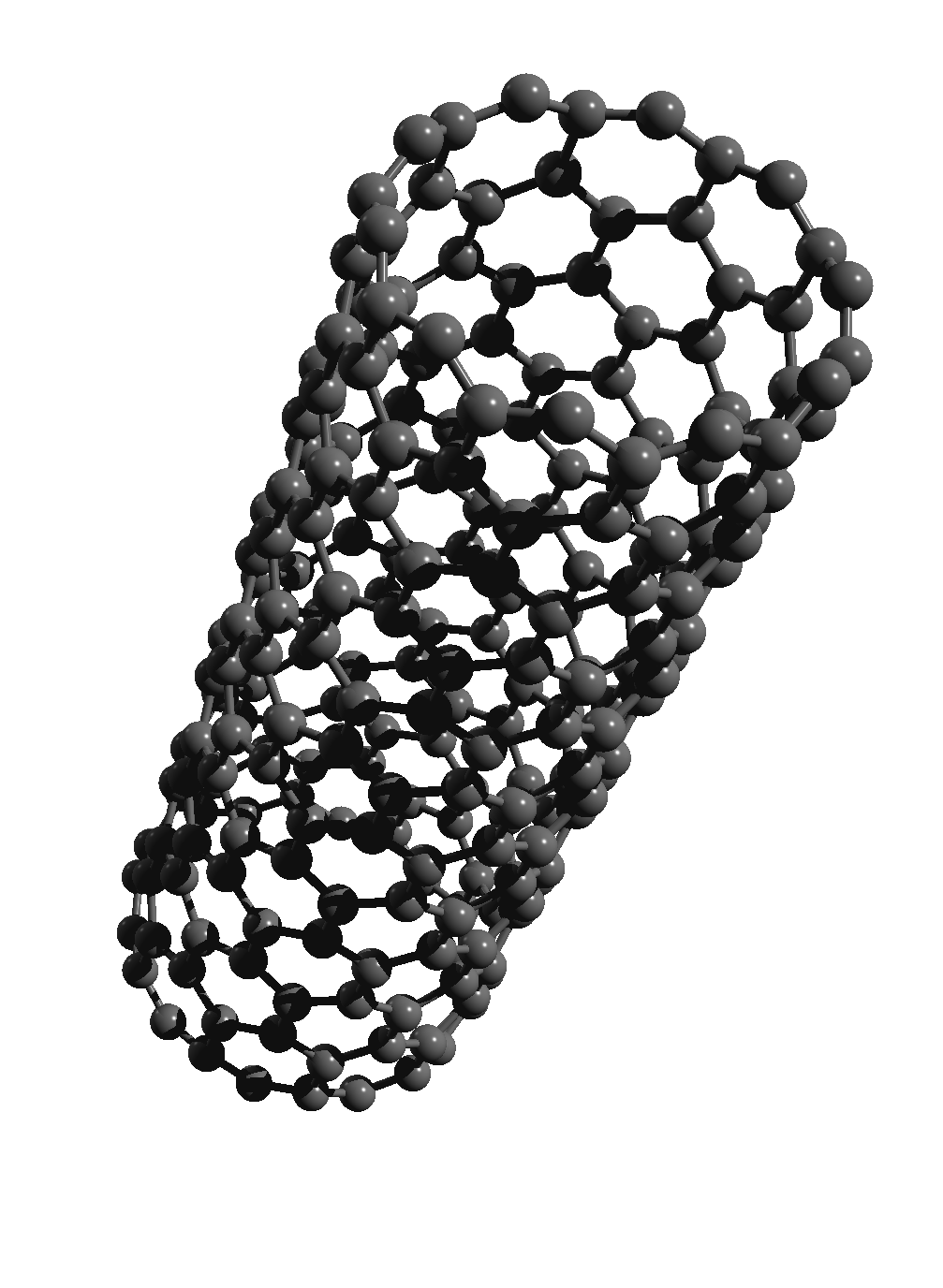}
\qquad
\includegraphics[width=50mm]{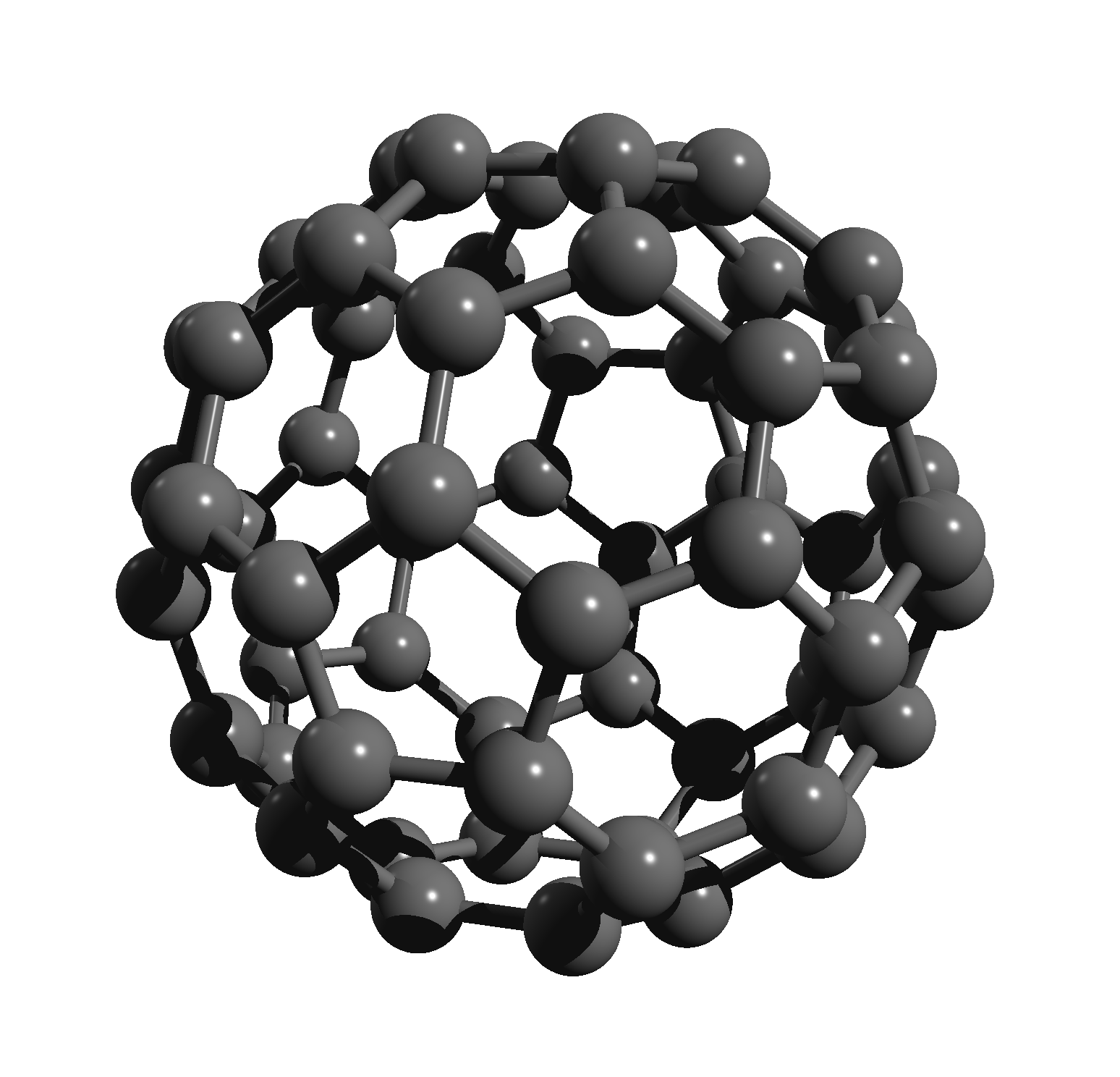}
\qquad
\includegraphics[width=45mm]{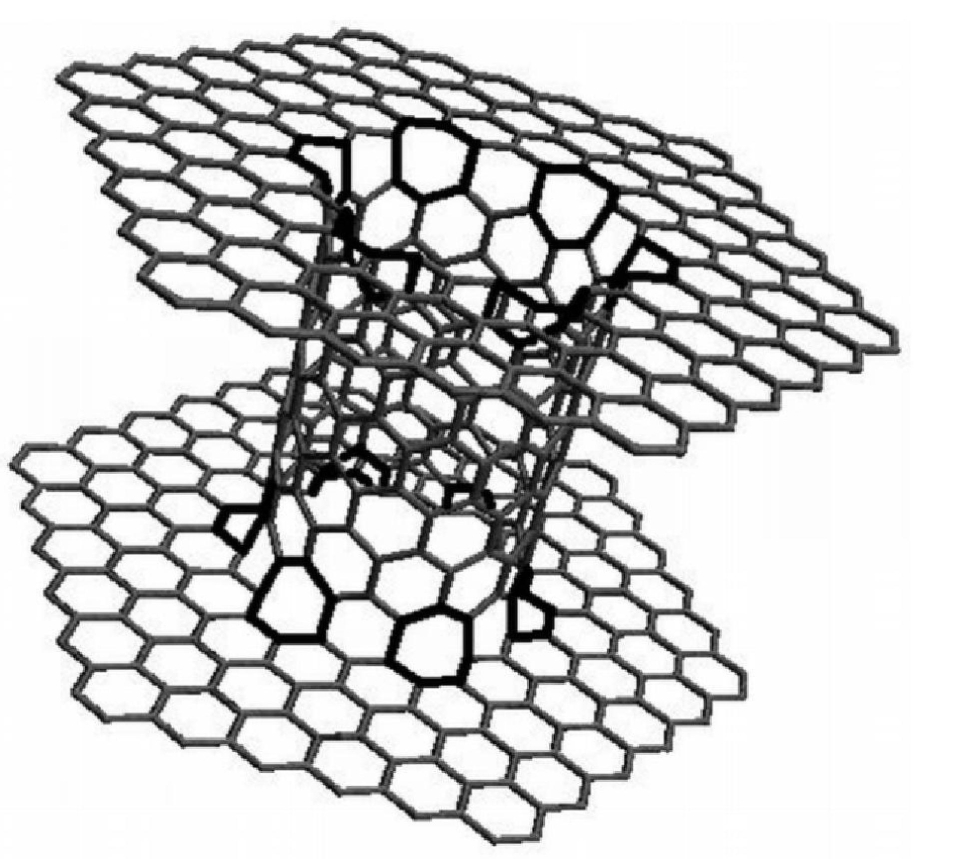}
\caption{Different kinds of graphene nanostructures: nanotube (left), fullerene (middle), wormhole (right).}\label{FG2}
\end{figure}

One more defect can be added and a nanotube is created. This nanostructure can be considered closed as well as opened, i.e. without the cap which contains the pentagonal defects. The second case is more common (Figure \ref{FG2}, left part). The number of the defects can be increased up to 12 and in this way, a completely closed, spherical nanostructure arises (fullerene - Figure \ref{FG2}, middle part).

\begin{figure}[htbp]
\includegraphics[width=40mm]{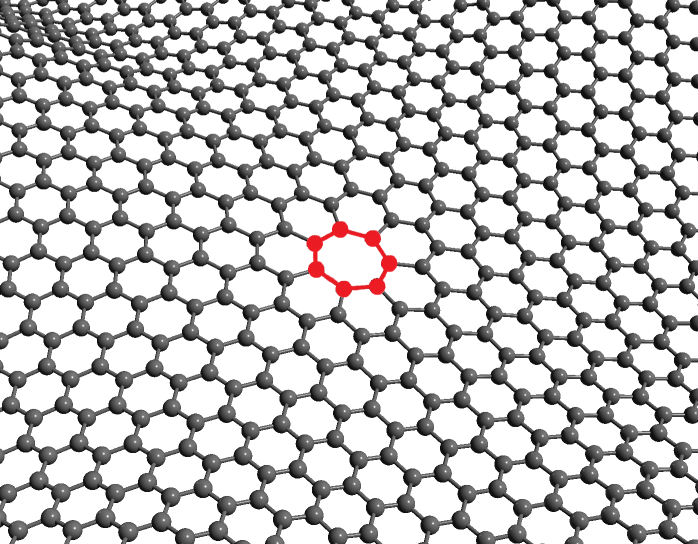}
\caption{Hexagonal lattice disclinated by 1 heptagonal defect.}\label{1hept}
\end{figure}

Analogical manipulations with the graphene lattice can be made by the supply of the heptagonal defects (Figure \ref{1hept}). For the case of 12 heptagonal defects, if they are placed appropriately, the wormhole structure is created (Figure \ref{FG2}, right part).

\begin{figure}[htbp]
\includegraphics[width=52mm]{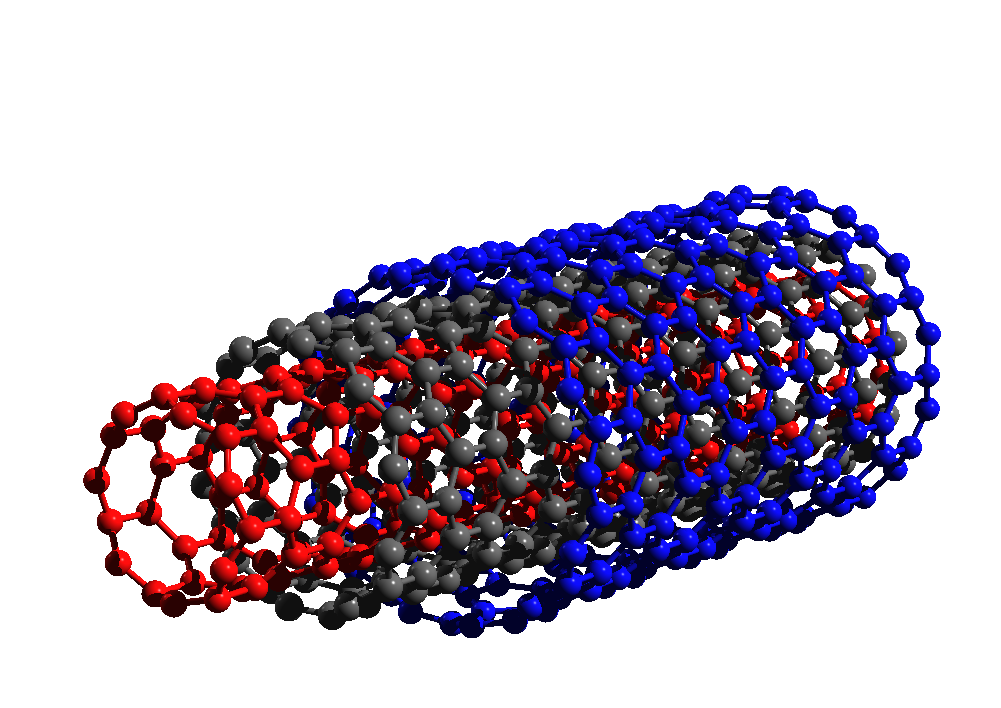}
\qquad
\includegraphics[width=52mm]{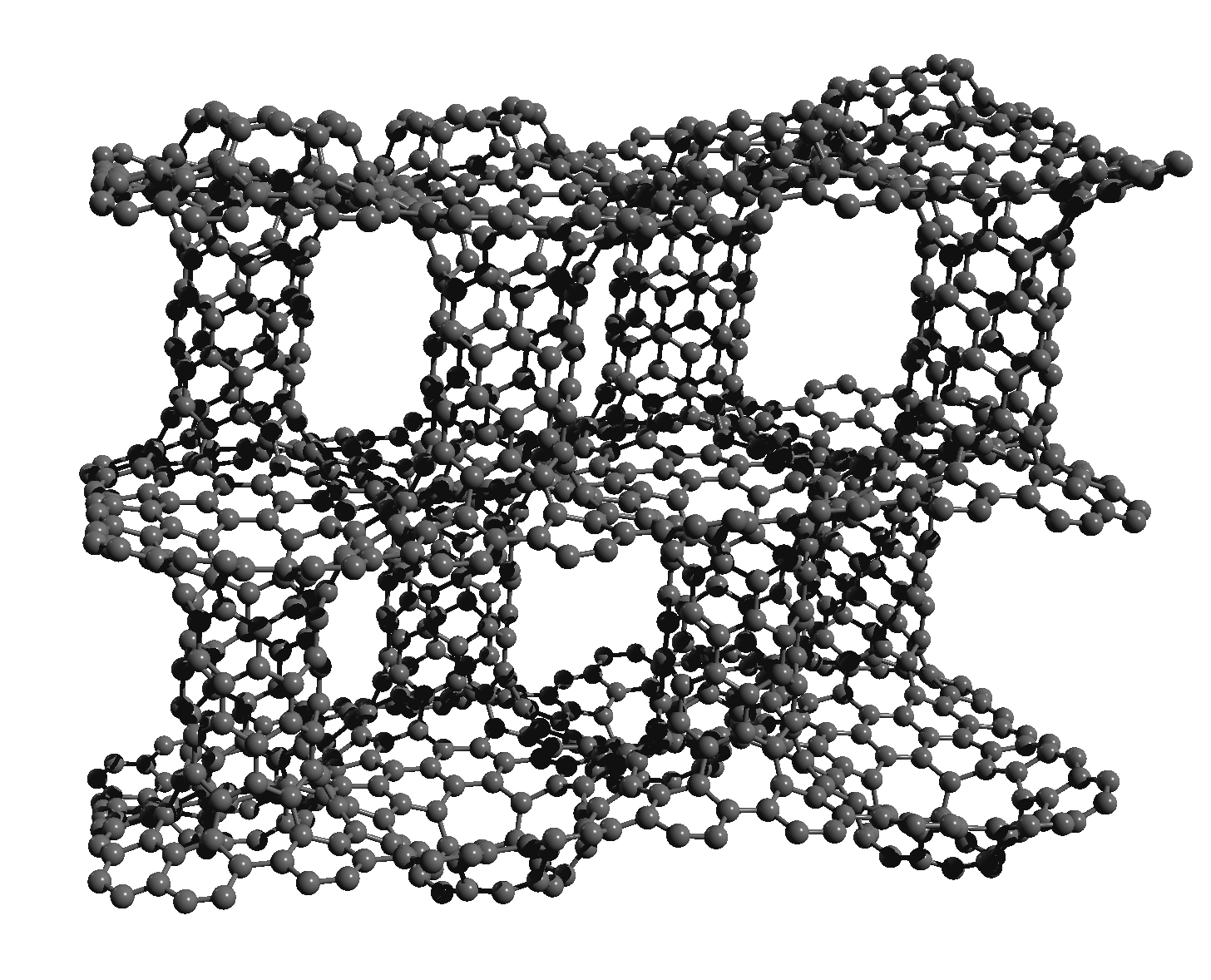}
\qquad
\includegraphics[width=52mm]{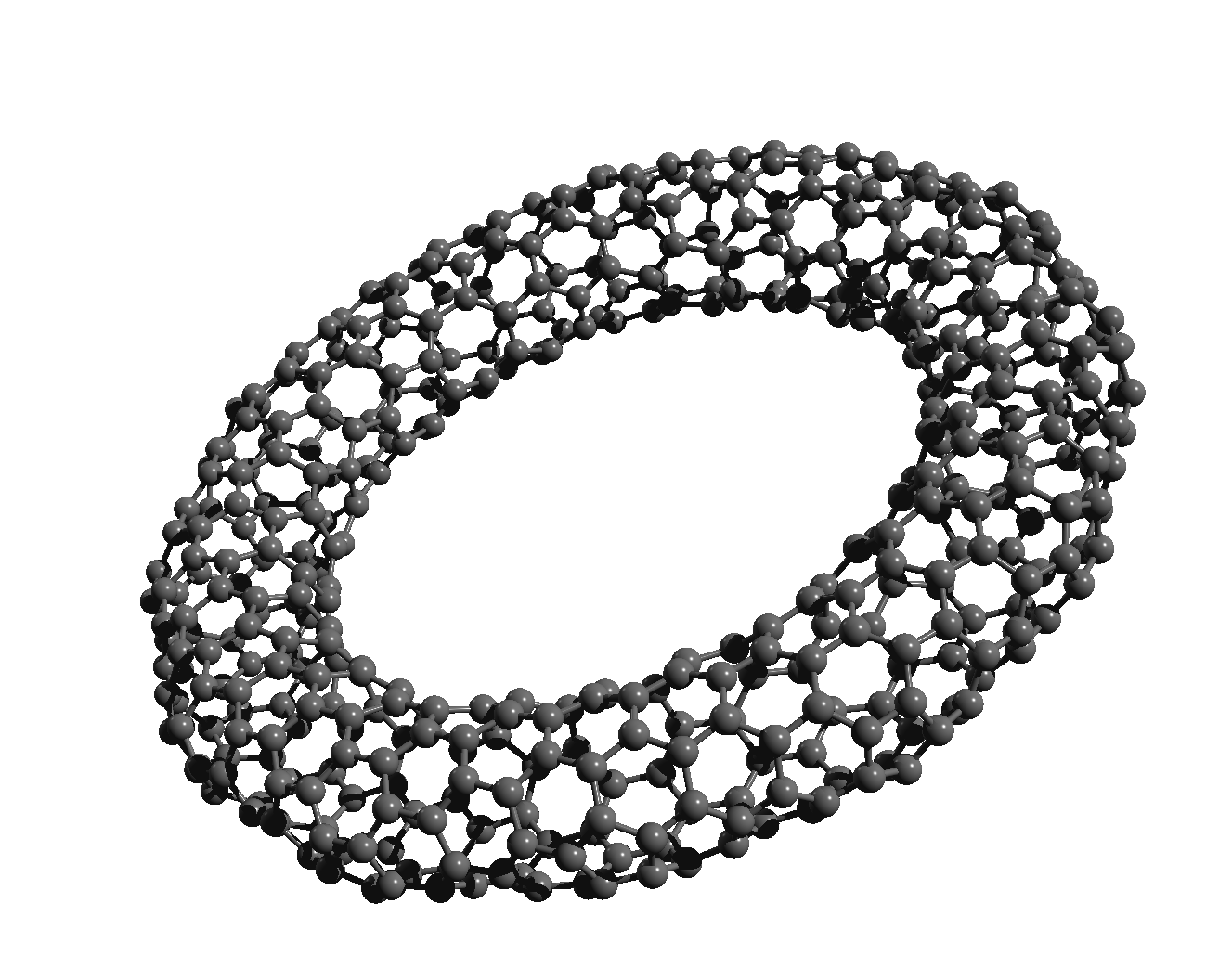}
\caption{Less common forms of the graphene nanostructures: triple-walled nanotube (left), pillared graphene (middle), nanotoroid (right).}\label{exot}
\end{figure}

A lot of other variants of the graphitic nanostructures can be created using different combinations of the pentagonal and the heptagonal defects. Some of them are presented in Figure \ref{exot}.

We investigate the electronic properties of several kinds of the carbon nanostructures. After the explanation of the computational methods, we demonstrate how to utilize these methods for the purpose of the investigation of graphene and some simple forms of the nanostructures - different kinds of nanoribbons and their modifications. Then, we will concentrate on the calculation of the properties of more complicated forms - the graphitic nanocone \cite{soc, nc2, boundaryNC} and the graphitic wormhole \cite{herrero1, analog, wormhole}. In the first case, we consider the influence of the additional effects like the spin--orbit coupling and the boundary effects coming from the finite size and from the extreme curvature of the surface geometry in the tip. In the second case, we investigate the effects which arise in the place of the wormhole bridge. Here, 2 additional effects appear: first, the spin--orbit coupling arising in the connecting nanotube and second, the increase of the electron mass due to relativistic effects coming from the extreme curvature of the surface geometry. As the result, the chiral massive electrons should be observed.\\

\section{Computational formalism}\

The electronic structure can be characterized by the density of states ($DOS$) - the number
of the electronic states per the unit interval of energies. This quantity can be used as the measure of the density of the electrons and generally we can say that the higher value of $DOS$, the higher conductivity. With the help of $DOS$, the electric field can be calculated as well. Besides $DOS$, one more quantity is defined - the local density of states ($LDOS$). It is $DOS$ related to the unit area of the molecular surface or to the unit area of the surface in the space of the wave vector $\vec{k}$. Then, the quantities depend on the variables as follows:
\begin{equation}DOS=DOS(E),\hspace{1cm}LDOS=LDOS(E,\vec{r}),\hspace{5mm}{\rm resp.}\hspace{5mm}LDOS=LDOS(E,\vec{k}).\end{equation}

2 methods are used for the calcuation of $LDOS$. The first one is used for the periodical structures with the planar geometry (plain graphene, nanotubes, nanoribbons, etc.), the second one is used for the structures which are aperiodical or which have the curved geometry (fullerene, nanocone, wormhole, nanotoroid etc.). We outline here the base of these methods. Both methods start on solving the Schr\"{o}dinger equation for the electron bounded on the molecular surface
\begin{equation}\label{SchrEq}\hat{H}\psi=E\psi.\end{equation}
Here,
\begin{equation}\label{HamUV}\hat{H}=\frac{\hat{k}^2}{2m}+\hat{V}(\vec{r})+\hat{U}(\vec{r}),\end{equation}
$\hat{V}$ representing the potential of the periodic crystal, $\hat{U}$ representing the external potential which is responsible for the curvature.\\

\subsection{Periodical structures with planar geometry}\

\begin{figure}[htbp]
\centering
\includegraphics[width=50mm]{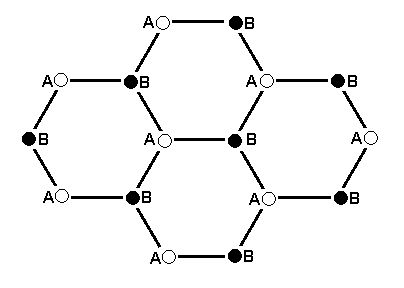}
\hspace{15mm}
\includegraphics[width=60mm]{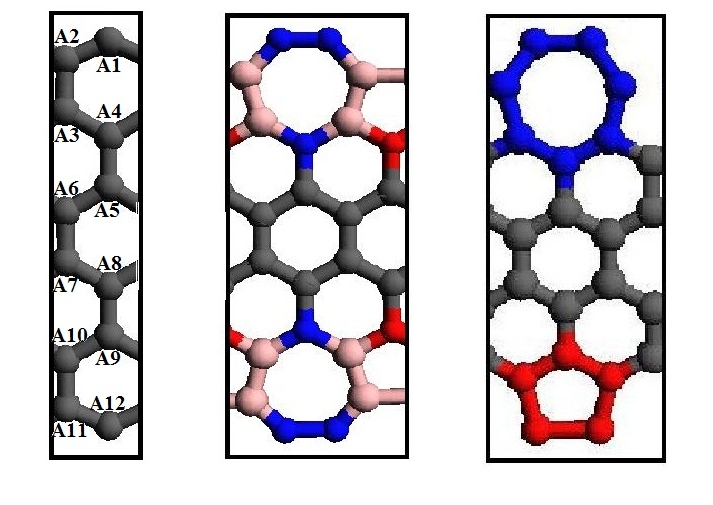}
\caption{Unit cells for different periodical structures: graphene (left), different kinds of nanoribbons (right). The graphene structure is considered to be infinite in 2 dimensions, the nanoribbons are considered to have final width and the second size is infinite as well.}
\label{unit}
\end{figure}

For the periodical structures, the external potential in Eq. (\ref{HamUV}) is zero, and the carbon lattice can be divided into several sublattices, each containing equivalent atomic sites. The sites corresponding to the different sublattices we can denote $A, B, C,...$ or $A_1, A_2, A_3,...$ We can find a unit cell - the smallest possible cell in the structure which contains all possible inequivalent atomic sites (Figure \ref{unit}).

\begin{figure}[htbp]
\centering
\includegraphics[width=152mm]{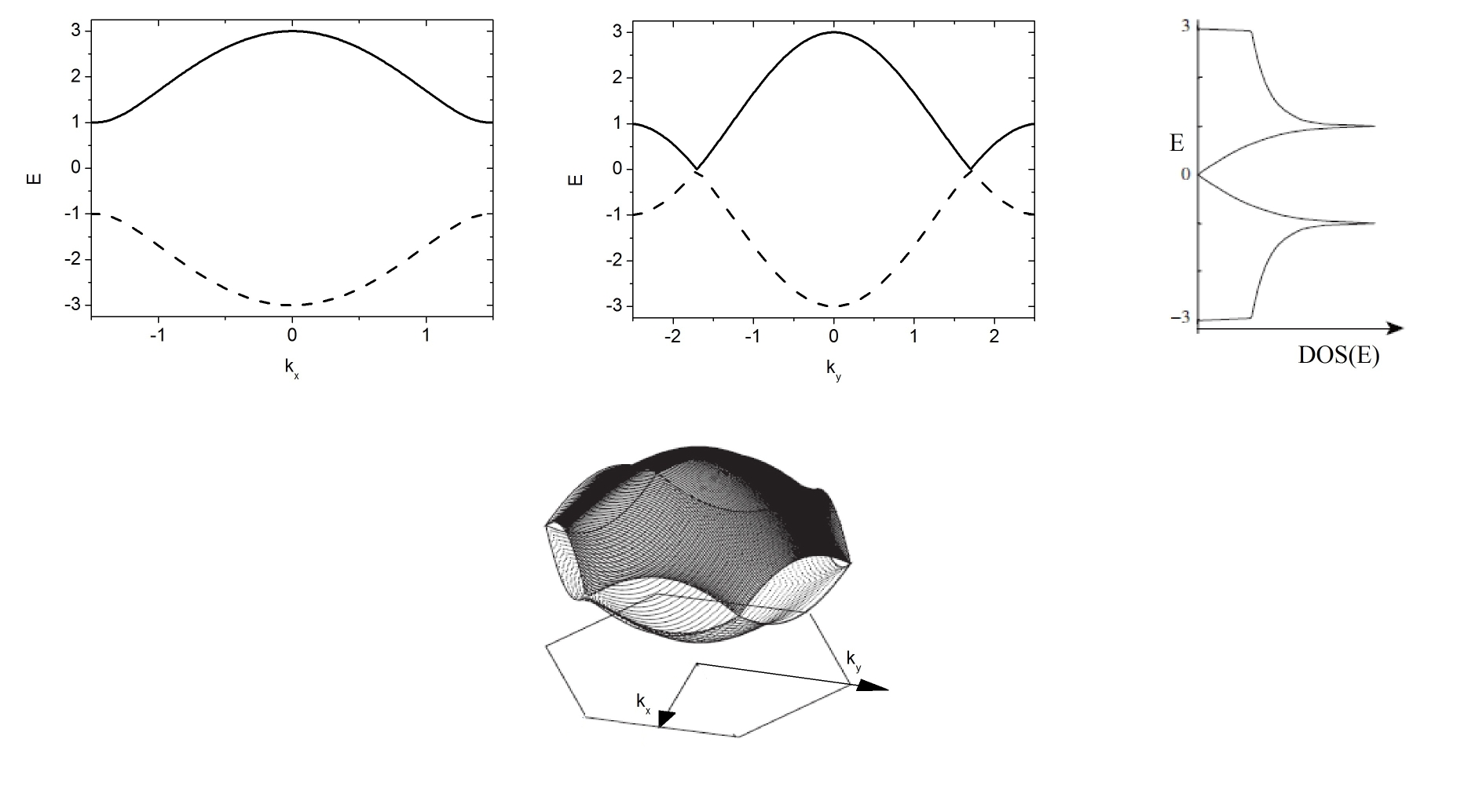}
\caption{The electronic spectrum for $k_y=0$ (left) and $k_x=0$ (middle) and the density of states (right) of graphene. The solid and dashed lines correspond to the positive and the negative energy values, respectively. In the bottom, we see the form of the electronic spectrum for an arbitrary value of the wave vector.}
\label{elspectrum}
\end{figure}

In the case of graphene, the wave function which solves Eq. (\ref{SchrEq})can be expressed as \cite{wallace, slonczewski}
\begin{equation}\label{solpsi}\psi=C_A\psi_A+C_B\psi_B,\end{equation}
where the components $\psi_A, \psi_B$ correspond to the particular sublattices. In the tight-binding approximation, we postulate the
solution in the form
\begin{equation}\label{solBloch}\psi_{A(B)}=\sum\limits_{A(B)}\exp[{\rm i}\vec{k}\cdot\vec{r}_{A(B)}]X(\vec{r}-\vec{r}_{A(B)}),\end{equation}
where $X(\vec{r})$ is the atomic orbital function. The overlap is zero, i.e.
\begin{equation}\int X^*(\vec{r}-\vec{r}_{A})X(\vec{r}-\vec{r}_{B}){\rm d}\vec{r}=0.\end{equation}
By the substitution of the solution in Eq. (\ref{solBloch}) into the Schr\"{o}dinger equation (Eq. (\ref{SchrEq})), multiplying it by $\psi^{\dagger}$ and making the integration over $\vec{r}$, we create the expressions
\begin{equation}H_{ab}=\int\psi^{*}_aH\psi_b{\rm d}\vec{r},\hspace{5mm}
S=\int\psi^{*}_A\psi_A{\rm d}\vec{r}=\int\psi^{*}_B\psi_B{\rm d}\vec{r},\hspace{5mm}a,b\equiv A,B.\end{equation}
If we suppose that the functions $X$ are normalized so that $\int X^*(\vec{r}-\vec{r}_{A(B)})X(\vec{r}-\vec{r}_{A(B)}){\rm d}\vec{r}=1$, then $S$ gives the number of the unit cells in the nanostructure. Now, the Schr\"{o}dinger equation is transformed into the matrix form
\begin{equation}\label{matrixeq}\left(\begin{array}{cc}
H_{AA} & H_{AB}\\ H_{BA} & H_{BB}
\end{array}\right)\left(\begin{array}{c}C_A\\C_B
\end{array}\right)=ES\left(\begin{array}{c}C_A\\C_B
\end{array}\right).
\end{equation}
The eigenvalues of the matrix in this equation are the energy eigenvalues and they create the electronic spectrum. For this purpose, first we need to determine the values of the matrix elements $H_{ab}$. From their definition follows
\begin{equation}H_{ab}=\int\psi_a^*H\psi_b{\rm d}\vec{r}=\sum\limits_{a,b}\exp\left[-{\rm i}\vec{k}\cdot(\vec{r}_{a}-\vec{r}_{b})\right]\int X^*(\vec{r}-\vec{r}_{a})HX(\vec{r}-\vec{r}_{b}){\rm d}\vec{r}=\sum\limits_{a,b}\exp\left[-{\rm i}\vec{k}\cdot(\vec{r}_{a}-\vec{r}_{b})\right]\gamma_{ab},\end{equation}
where $\gamma_{ab}$ denotes the corresponding hopping integral. The eigenvalues are labeled by $\vec{k}$ and in the nearest-neighbor approximation, they can be expressed as
\begin{equation}E(\vec{k})=\pm\gamma_0\sqrt{1+4\cos^2\frac{k_ya}{2}+4\cos\frac{k_ya}{2}\cos\frac{k_x a
\sqrt{3}}{2}},\end{equation}
where $\gamma_0=\gamma_{AA}=\gamma_{BB}$ is the hopping integral for the nearest neighboring atoms and $a=2.46$A is the distance between the next-nearest atomic neighbors. $DOS$ and $LDOS$ are then defined as
\begin{equation}DOS(E)=-\frac{1}{\pi}\lim\limits_{\omega\rightarrow0}\,{\rm Im}\int\frac{{\rm d}\vec{k}}{E-E(k)+{\rm i}\omega},\hspace{1cm}LDOS(E,k)=\delta(E-E(k)).\end{equation}
The corresponding graphs of electronic spectrum and $DOS$ are given in Figure \ref{elspectrum}. We see that for zero energy, the density of states has zero value. This property is typical for the semimetallic nanostructures. For the metallic nanostructures, a peak appears for zero energy. In the first case, a gap is present around zero in the electronic spectrum. Its width can be influenced by the additional defects in the hexagonal structure or by the chemical admixtures and in this way, a material with the pre-defined properties can be synthesized. In the second case, the gap around zero energy is absent.\\

In a similar way, but with a more complicated structure of the wave function in Eq. (\ref{solpsi}) and for a larger size of the matrix in Eq. (\ref{matrixeq}), the electronic spectrum and $DOS$ can be found for other nanostructures like the nanoribbons in the right side of Figure \ref{unit} \cite{stonewales}.

\begin{figure}[htbp]
\includegraphics[width=130mm]{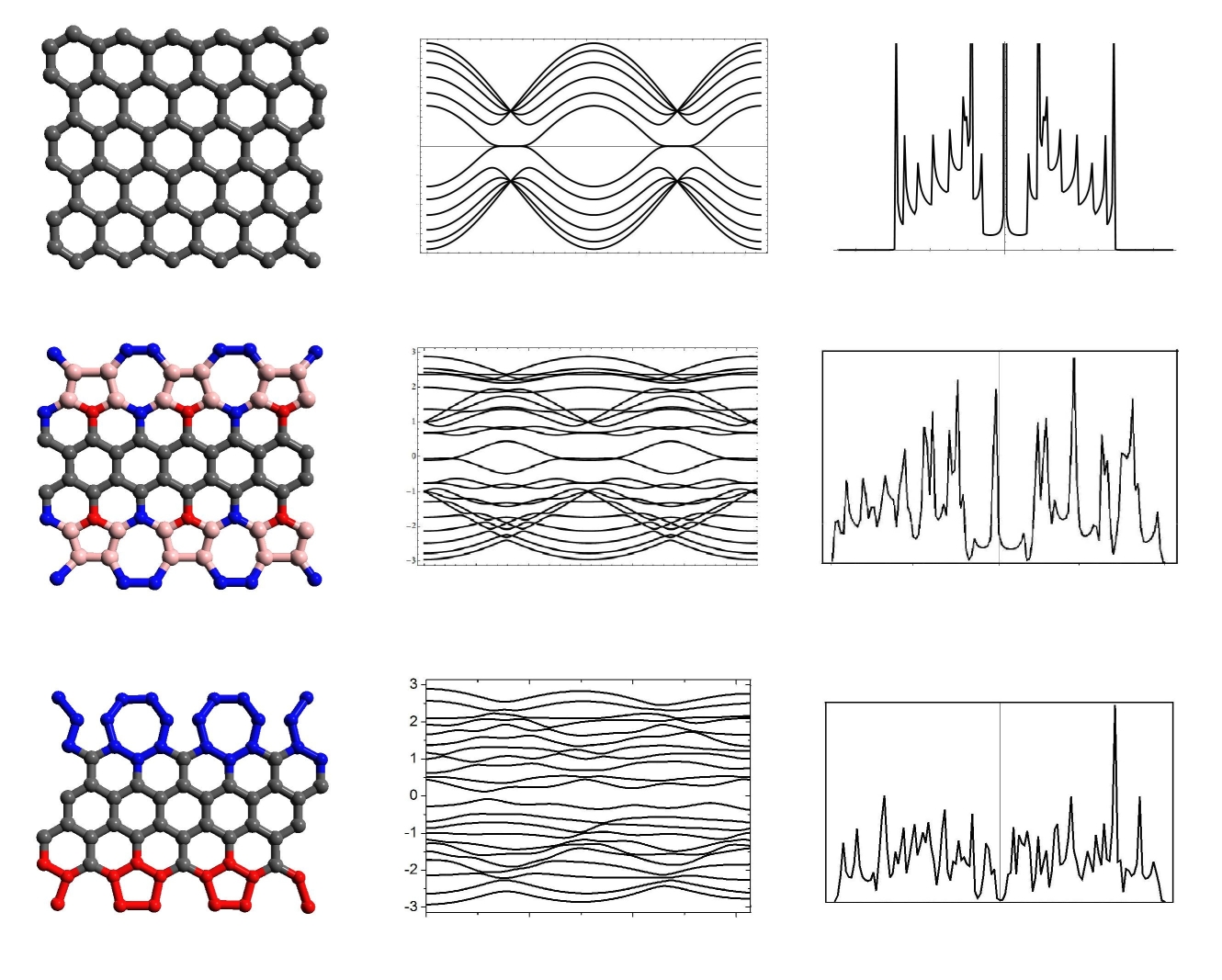}
\caption{Electronic spectrum and $DOS$ of different kinds of nanoribbons.}
\label{nanoribbons}
\end{figure}

The results we see in Figure \ref{nanoribbons}. In the left part, the shape of the segment of the concrete nanoribbon is present. The plot of the electronic spectrum and $DOS$ are given in the middle and in the right part, respectively. The direction of the wave vector $\vec{k}$ is considered longitudinal.

The upper part corresponds to the nanoribbon with zigzag edges \cite{wakab} and with 12 atomic sites in the unit cell (see Figure \ref{unit}). That is why the size of the appropriate matrix in Eq. (\ref{matrixeq}) would be  $12\times 12$ and its spectrum contains 12 eigenvalues. This corresponds to 12 lines in the graph of the electronic spectrum. The graph of $DOS$ shows a zero energy peak which signalizes the metallicity of this kind of nanostructure. It is a typical property for the zigzag nanoribbons unlikely the armchair nanoribbons \cite{wakab}.

\begin{figure}[htbp]
\includegraphics[width=65mm]{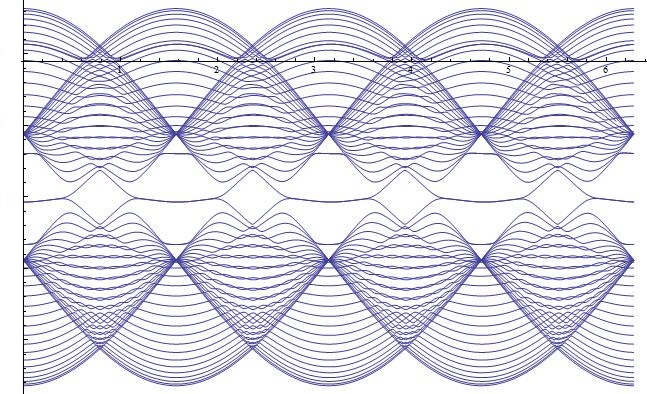}
\caption{Electronic spectrum for extra wide nanoribbon with reconstructed edges.}\label{SWlarge}
\end{figure}

The middle and the bottom part correspond to some modifications of the previous form - the nanoribbon with the reconstructed edges. This causes the enlargement of the unit cell (Figure \ref{unit}) and, consequently, more complicated structure of the electronic spectrum. The metallic properties depend on the concrete kind of the modification: for the nanostructure in the middle part, the zero energy peak in $DOS$ is preserved, while it disappears for the nanostructure in the bottom part. Furthermore, in the first case, the electronic spectrum gets more complicated structure - the number of the Dirac points, where the lines are crossing, is doubled. This feature remains and strengthens for the larger width: in Figure \ref{SWlarge}, the form of the electronic spectrum is depicted for the same kind of the nanostructure its width is 3 times larger.\\

\subsection{\label{AS}Structures with curved geometry}\

For the curved structures, the non-trivial geometry is described by the external potential $U(\vec{r})$ in Eq. (\ref{HamUV}). Because of the non-periodicity, the eigenvalues cannot be labeled by the wave vector $\vec{k}$. Nevertheless, the solution of the Schr\"{o}dinger equation (Eq. (\ref{SchrEq})) can be expressed with the help of the solutions for the previous case as
\begin{equation}\label{k_labeled}\Psi(\vec{r})=\int{\rm d}\vec{k}\,\psi_{\vec{k}}(\vec{r}),\end{equation}
so, labeling by the wave vector will still play a key role in the following procedure.

For the purpose of the calculations, we express the wave function which solves Eq. (\ref{SchrEq}) in the case of zero external potential in the form of the so-called Luttinger-Kohn base \cite{mele}:
\begin{equation}\label{lutkon}\psi_{\vec{k}}(\vec{r})=f_A(\vec{\kappa})e^{{\rm i}\vec{\kappa}\cdot\vec{r}}\psi_A(\vec{K},\vec{r})+f_B(\vec{\kappa})e^{{\rm i}\vec{\kappa}\cdot\vec{r}}\psi_B(\vec{K},\vec{r}),\end{equation}
where $\vec{\kappa}=\vec{k}-\vec{K}$, $\vec{K}$ being the Dirac point and $E(\vec{k})$ is the appropriate eigenvalue for the zero external potential. After the substitution of this expression into Eq. (\ref{k_labeled}), we get
\begin{equation}\label{nonzeroU}\Psi(\vec{r})=\int{\rm d}\vec{\kappa}\left(f_A(\vec{\kappa})e^{{\rm i}\vec{\kappa}\vec{r}}\psi_A(\vec{K},\vec{r})+f_B(\vec{\kappa})e^{{\rm i}\vec{\kappa}\vec{r}}\psi_B(\vec{K},\vec{r})\right),\end{equation}
so, the resulting form of the Schr\"{o}dinger equation is
\begin{equation}\label{extschreq}\hat{H}\Psi(\vec{r})=\left(\frac{\hat{k}^2}{2m}+\hat{V}(\vec{r})+\hat{U}(\vec{r})\right)\Psi(\vec{r})
=E\Psi(\vec{r}).\end{equation}
Now, we substitute Eq. (\ref{lutkon}) into Eq. (\ref{SchrEq}) for the case of zero external potential. After some manipulations and the substitution of the result into Eq. (\ref{nonzeroU}), we substitute the resulting wave function into Eq. (\ref{extschreq}) and get \cite{callaway}
{\scriptsize\[\int e^{i(\vec{k}-\vec{q})\cdot\vec{r}}
u^*_j(\vec{K},\vec{r})\left(E(\vec{K})+\frac{\hbar}{m}\vec{\kappa}\cdot \vec{p}+\frac{\hbar^2}{2m}(\vec{k}^2-\vec{K}^2)+U(\vec{r})\right)
\left(f_A(\vec{\kappa})u_{A}(\vec{K},\vec{r})+
f_B(\vec{\kappa})u_{B}(\vec{K},\vec{r})\right){\rm d}\vec{\kappa}=\]}
\begin{equation}\label{EMrepr}=E\int e^{i(\vec{k}-\vec{q})\cdot\vec{r}}
u^*_j(\vec{K},\vec{r})\left(f_A(\vec{\kappa})u_{A}(\vec{K},\vec{r})+
f_B(\vec{\kappa})u_{B}(\vec{K},\vec{r})\right){\rm d}\vec{\kappa}.
\end{equation}
Here, we used the substitution
\begin{equation}\label{subst}\psi_{A(B)}(\vec{K},\vec{r})=e^{i\vec{K}\cdot\vec{r}}
u_{A(B)}(\vec{K},\vec{r}),\end{equation}
where $u_A, u_B$ are the functions which are periodical in the crystal lattice. After making the appropriate integrations over $\vec{r}$ and $\vec{q}$, the resulting integrand has the form
\[\left(E(\vec{K})+\frac{\hbar^2}{2m}(\vec{k}^2-\vec{K}^2)
-E\right)f_i(\vec{\kappa})+\frac{\hbar}{m}\vec{\kappa}\cdot\left(\vec{p_{iA}}(\vec{K})
f_A(\vec{\kappa})+\vec{p_{iB}}(\vec{K})f_B(\vec{\kappa})\right)+\]
\begin{equation}+\int{\rm d}\vec{q}\left(\langle i,\vec{k}|U|A,\vec{q}\rangle f_A(\vec{\kappa})+
\langle i,\vec{k}|U|B,\vec{q}\rangle f_B(\vec{\kappa})\rangle\right)=0,\end{equation}
where $i=A,B$,
\begin{equation}\label{momint}\vec{p_{ab}}=\int\psi_a^*\vec{p}\psi_b{\rm d}\vec{r}\end{equation}
and
\begin{equation}\langle i,\vec{k}|U(\vec{r})|j,\vec{q}\rangle=
\int e^{i(\vec{q}-\vec{k})\cdot\vec{r}}u^*_i(\vec{K},\vec{r})
U(\vec{r})u_j(\vec{K},\vec{r}){\rm d}\vec{r}.\end{equation}
Now, using the Fourier transform and by the diagonalization of the corresponding matrix \cite{mele}, we finally get
\begin{equation}\frac{P}{i}\left(\begin{array}{cc}
0 & \frac{\partial}{\partial x}-i\frac{\partial}{\partial y}\\
\frac{\partial}{\partial x}+i\frac{\partial}{\partial y} &
0\end{array}\right)\left(\begin{array}{c}F_1(\vec{r})\\F_2(\vec{r})
\end{array}\right)=[E-U_2(\vec{r})]\left(\begin{array}{c}F_1(\vec{r})\\F_2(\vec{r})
\end{array}\right),\end{equation}
where $P$ is the averaged value of the expression $\frac{E(\vec{\kappa})}{|\vec{\kappa}|}$, $E(\vec{\kappa})=E(\vec{k})-E(\vec{K})$ and $U_2(\vec{r})$ comes from the transformation of the potential $U(\vec{r})$ \cite{callaway}. The last expression means that Eq. (\ref{nonzeroU}) can be expressed as
\begin{equation}\Psi(\vec{r})=F_1(\vec{r})
\psi_A(\vec{K},\vec{r})+
F_2(\vec{r})\psi_B(\vec{K},\vec{r}).\end{equation}
We can understand it as a 2-component wave function
\begin{equation}\Psi^T(\vec{r})=(F_1(\vec{r}),F_2(\vec{r}))\end{equation}
and then
\begin{equation}\label{Dirac}\left(\frac{P}{i}
\vec{\sigma}\cdot\vec{\nabla}+U_2(\vec{r})-E\right)\Psi(\vec{r})=0.\end{equation}
Here, $\vec{\sigma}$ has the same form  as the Pauli matrices for the indices $1,2$:
\begin{equation}\label{Pauli}\sigma_1=\left(\begin{matrix}0 & 1\\1 & 0\end{matrix}\right),\hspace{1cm}
\sigma_2=\left(\begin{matrix}0 & -i\\i & 0\end{matrix}\right).
\end{equation}
But this is algebraically identical to a 2-dimensional Dirac equation for the massless fermion, where
the 2 components correspond to the graphene sublattice A or B, respectively.\\

In the practical calculations, a suitable choice of the coordinates is useful. In our case, we will suppose the rotational symmetry. Then, we perform the transformation of the coordinates: $(x,y,z)\,\rightarrow\,(\xi,\varphi)$, where $\varphi$ is the angular coordinate. Then, the resulting equation will have the form \cite{kochetov, smotlacha}
\begin{equation}\label{DirEq}{\rm i}\sigma^{\alpha}e_{\alpha}^{\mu}[\partial_{\mu}+\Omega_{\mu}-{\rm i}a_{\mu}-{\rm i}a_{\mu}^W-{\rm i}A_{\mu}]\Psi=E\Psi.\end{equation}
The transition from Eq. (\ref{Dirac}) to this expression comes from the change of the coordinates and from the given choice how to represent the external potential $U_2(\vec{r})$. The meaning of the particular terms is following: $e_{\alpha}^{\mu}$, the zweibein, is connected with metric and using the corresponding tensor, it can be defined as
\begin{equation}g_{\mu\nu}(x)=e^{\alpha}_{\mu}(x)e^{\beta}_{\nu}(x)\eta_{\alpha\beta}.\end{equation}
Here, $\eta_{\alpha\beta}$ is the metric of the plain space without curvature. Next term, $\Omega_{\mu}$, which is the spin connection in the spinor representation, is defined as $\Omega_{\mu}=\frac{1}{8}\omega^{\alpha\beta}_{\mu}[\sigma_{\alpha},\sigma_{\beta}]$. Here, $\omega_{\mu}$ is a more usual form of the spin connection. For its definition, we have to stress that the rotational symmetry is supposed. Then, it has the form
\begin{equation}\omega^{12}_{\varphi}=-\omega^{21}_{\varphi}=1-\frac{\partial_{\xi}\sqrt{g_{\varphi\varphi}}}
{\sqrt{g_{\xi\xi}}}=2\omega,\hspace{1cm}\omega^{12}_{\xi}=\omega^{21}_{\xi}=0.\end{equation}
It remains to explain the sense of the gauge fields $a_{\mu}, a_{\mu}^W, A_{\mu}$. First 2 of them ensure the circular periodicity. Their form is
\begin{equation}a_{\varphi}=N/4,\hspace{1cm}a_{\varphi}^W=-\frac{1}{3}(2m+n),\end{equation}
where $N$ is the number of defects and $(n,m)$ is the chiral vector. The last term, $A_{\mu}$, represents one possible additional effect - the magnetic field.

The rotational symmetry enables to find the solution of Eq. (\ref{DirEq}) with the help of the substitution
\begin{equation}\label{rotsubs}\Psi^T=\left(\begin{array}{c}F_1 \\ F_2\end{array}\right)=\frac{1}{\sqrt[4]{g_{\varphi\varphi}}}\left(\begin{array}{c}u_j(\xi)e^{i\varphi
j}\\ v_j(\xi)e^{i\varphi(j+1)}\\\end{array}\right),\hspace{1cm} j=0,\pm
1,...,\end{equation}
from which we get the system
\begin{equation}\label{Dsystem}\frac{\partial_{\xi}u_j}{\sqrt{g_{\xi\xi}}}-\frac{\widetilde{j}}
{\sqrt{g_{\varphi\varphi}}}u_j
=Ev,\hspace{1cm}-\frac{\partial_{\xi}v_j}{\sqrt{g_{\xi\xi}}}-\frac{\widetilde{j}}
{\sqrt{g_{\varphi\varphi}}}v_j
=Eu.\end{equation}
Here,
\begin{equation}\widetilde{j}=j+1/2-a_{\varphi}-a_{\varphi}^W-A_{\varphi}.\end{equation}
From the solution, $LDOS$ is defined as the square of the absolute value of the wave function. In this case, we get it as the sum of squares of the absolute values of its $\xi$-components:
\begin{equation}\label{LDfin}LDOS(E,\xi)=|u(E,\xi)|^2+|v(E,\xi)|^2.\end{equation}\\

\section{\label{propNC} Properties of the graphitic nanocone}\

\begin{figure}[htbp]
\begin{minipage}[b]{0.4\linewidth}
\centering
\includegraphics[width=55mm]{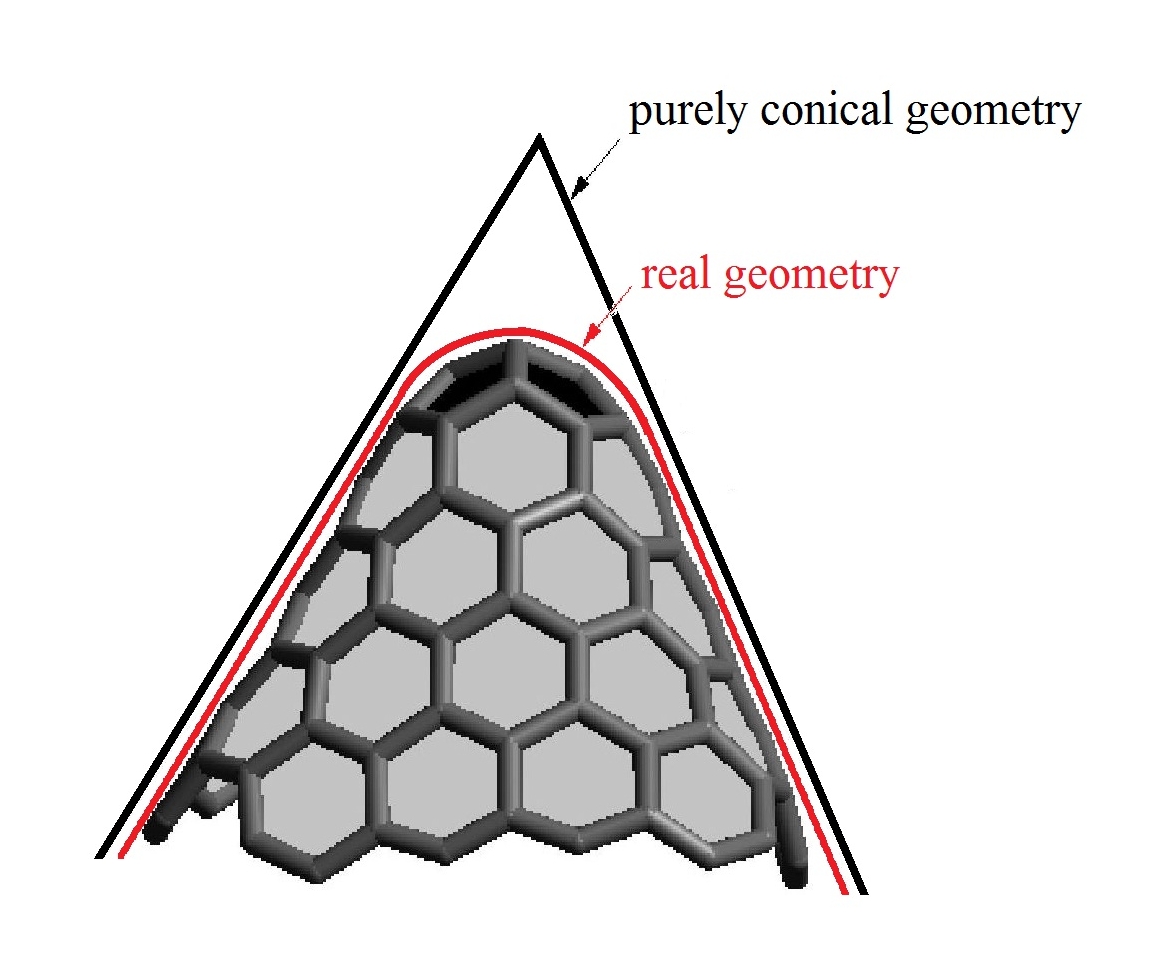}
\caption{The deviation of the geometry of the graphitic nanocone from the geometry of the real nanocone.}\label{geom}
\end{minipage}
\qquad\qquad
\begin{minipage}[b]{0.4\linewidth}
\centering
\includegraphics[width=35mm]{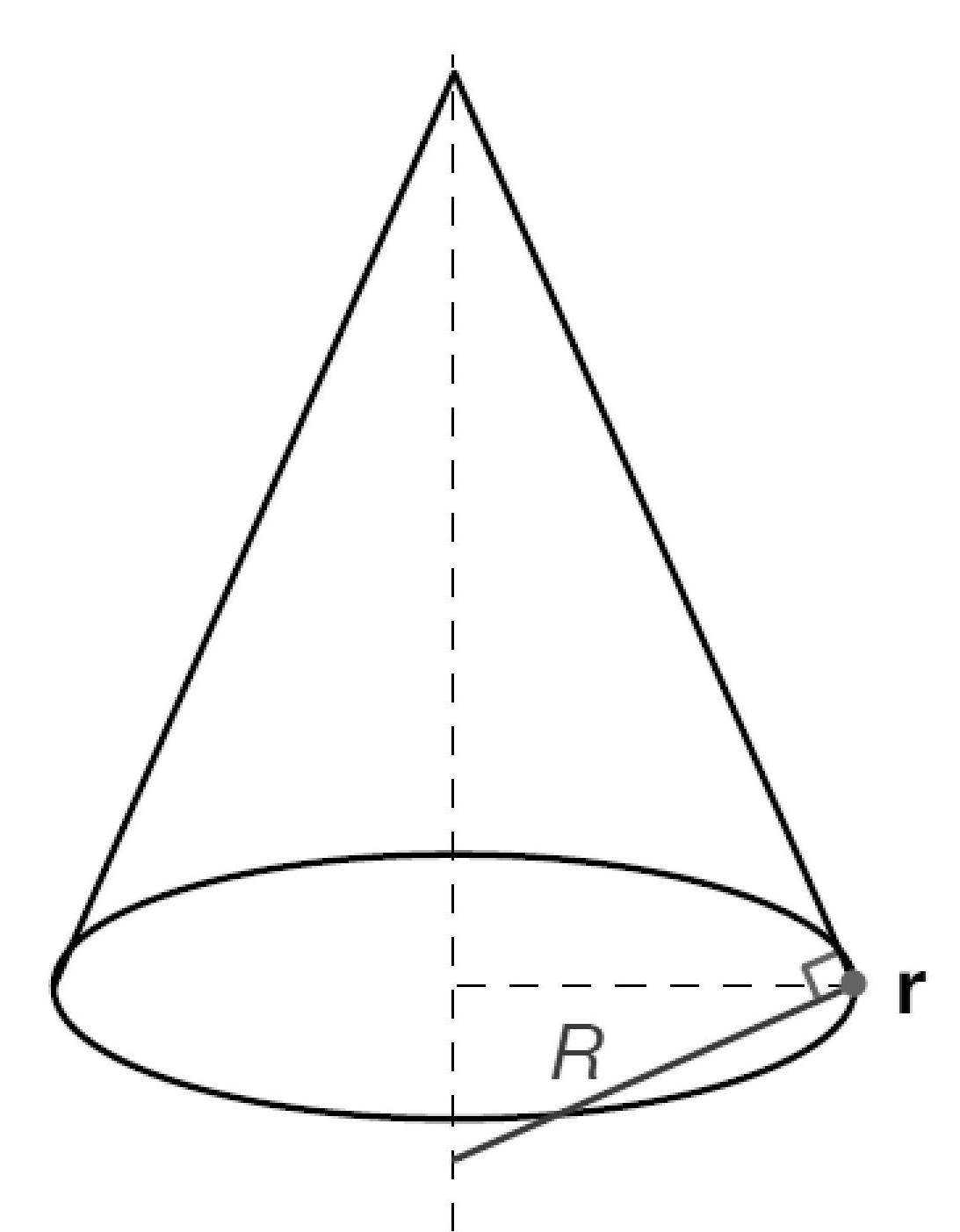}
\caption{The notation of the variables in the nanocone.}\label{konus}
\end{minipage}
\end{figure}

The graphitic nanocone is a nanostructure which can be created from the plain graphene by the insertion of the pentagonal defects into the hexagonal structure. The number of these defects can vary from 1 to 5. In this way, the conical tip arises and its smoothness and the vortex angle is given by the number of the defects and their placement. Then, the real geometry of the graphitic nanocone and the pure conical geometry are different (Figure \ref{geom}). The value of the vortex angle $\varphi$ for the purely conical geometry can be calculated as
\begin{equation}\sin\frac{\varphi}{2}=1-\frac{N}{6},\end{equation}
where $N$ is the number of the pentagonal defects in the conical tip.

The electronic structure of the graphitic nanocone with purely conical geometry and without any additional effects was investigated in \cite{sitenko}. There, the solution of Eq. (\ref{DirEq}) for this case is derived. Here, using the gauge field--theory approach, we introduce the results of the calculations in different approximations: the nanocone with purely conical geometry influenced by the spin--orbit coupling (SOC, \cite{soc}) and the same case with the additional effect of the Coulomb interaction coming from the charge placed into the conical tip \cite{boundaryNC}. The reason is following: it is one of the possibilities how to simulate the real geometry in the conical tip.\\

\subsection{Electronic structure influenced by the spin--orbit coupling}\

In the case of the purely conical structure, this form is assigned to the Hamiltonian in the Schr\"{o}dinger equation (Eq. (\ref{SchrEq})) \cite{sitenko}:
\begin{equation}\label{H0}\hat{H}_0=\left(\begin{array}{cc}H_1 & 0\\0 & H_{-1}\end{array}\right)\hspace{1cm}
\hat{H}_{0s}={\rm i}\hbar v_F\left\{\tau^y\partial_r-\tau^xr^{-1}\left[(1-\eta)^{-1}\left(s\partial_{\varphi}-\frac{3}{2}{\rm i}\eta\right)-
\frac{1}{2}\tau^z\right]\right\}.\end{equation}
In this equation, $v_F$ is the Fermi velocity, $s=\pm 1$ denotes the value of the $K$ spin, $\eta=N/6$, $\tau^x, \tau^y, \tau^z$ are the Pauli matrices - these matrices have nothing to do with SOC. The points on the surface are described by the coordinates $r$ and $\varphi$. The value of $r$ is given by the distance from the tip (see Figure \ref{konus}).

SOC is incorporated using the substitutions \cite{soc}
\begin{equation}\partial_r\,\rightarrow\,\partial_r-\frac{\delta\gamma'}{4\gamma R}\sigma_x(\vec{r}),
\hspace{1cm}{\rm i}\partial_{\varphi}\,\rightarrow\,{\rm
i}\partial_{\varphi}+s(1-\eta)A_y\sigma_y,\end{equation}
where $R=\frac{(1-\eta)r}{\sqrt{\eta(2-\eta)}}, A_y=s\frac{2\delta p}{(1-\eta)}\sqrt{\eta(2-\eta)}$. Here, in the analogy to the case of the nanotube, $R$ represents the curvature \cite{Ando} (the geometrical meaning is also presented in Figure \ref{konus}), $A_y$ is connected with the curvature of the carbon bonds. Next, $\gamma$ and $\gamma'$ represent the nearest neighbor and the next-nearest neighbor hopping integrals, respectively and $\sigma_{x}(\vec{r})=\sigma^{x}\cos\varphi-\sigma^{z}\sin\varphi$. Here, $\sigma^x, \sigma^y, \sigma^z$ are the Pauli matrices, but (unlikely the $\tau$ matrices) this time they are connected with SOC. The parameters $p$ and $\delta$ are connected with the hopping integrals and the atomic potential, respectively. Their closer explanation can be also found in \cite{Ando}. The following choice of their values was used: $\delta$ is of the order between $10^{-3}$ and $10^{-2}$, $\frac{\gamma'}{\gamma}\sim\frac{8}{3}$, $p\sim 0.1$.

After making the transformation $\hat{H}_s\rightarrow e^{{\rm i}\frac{\sigma_y}{2}\varphi}\hat{H}_se^{-{\rm i}\frac{\sigma_y}{2}\varphi}$ which transforms the coordinate frame into the local coordinate frame, the final form of the Hamiltonian is
\begin{equation}\label{HamSOC}\hat{H_{s}}=\hbar v_F\left(\begin{array}{cc}0 & \partial_r-{\rm i}\frac{1}{r}\xi_x\sigma_x(\vec{r})-\frac{{\rm i}s\partial_{\varphi}}{(1-\eta)r}-\frac{A_y}{r}\sigma_y-\frac{3\eta}{2(1-\eta)r}+\frac{1}{2r}\\
-\partial_r+{\rm i}\frac{1}{r}\xi_x\sigma_x(\vec{r})-\frac{{\rm i}s\partial_{\varphi}}{(1-\eta)r}-\frac{A_y}{r}\sigma_y-\frac{3\eta}{2(1-\eta)r}-\frac{1}{2r} & 0\end{array}\right).\end{equation}
It includes the strength of SOC through the parameters $\xi_x, \xi_y$:
\begin{equation}\xi_x=\frac{\delta\gamma'\sqrt{\eta(2-\eta)}}{4(1-\eta)\gamma},\hspace{1cm}                                \xi_y=A_{y}+\frac{1}{2(1-\eta)}.\end{equation}

Now, the equation
\begin{equation}\label{DirEq2}\hat{H}_s\psi(r,\varphi)=E\psi(r,\varphi)\end{equation}
will be solved for the calculation of $LDOS$. Similarly as in Eq. (\ref{rotsubs})), we can do the following factorization due to the rotational symmetry:
\begin{equation}\psi(r,\varphi)=e^{{\rm i}j\varphi}\left(\begin{array}{c}f_{j\uparrow}(r)\\
f_{j\downarrow}(r)\\g_{j\uparrow}(r)\\g_{j\downarrow}(r)\end{array}\right)\end{equation}
It changes the equation into the form
\begin{equation}\label{syst}\left(\begin{array}{cccc}0 & 0 & \partial_r+\frac{F}{r} & -\frac{\rm i}{r}C\\0 & 0 & -\frac{\rm i}{r}D &
\partial_r+\frac{F}{r}\\ -\partial_r+\frac{F-1}{r} & \frac{\rm i}{r}D & 0 & 0\\
\frac{\rm i}{r}C & -\partial_r+\frac{F-1}{r} & 0 & 0\end{array}\right)\left(\begin{array}{c}f_{j\uparrow}(r)\\
f_{j\downarrow}(r)\\g_{j\uparrow}(r)\\g_{j\downarrow}(r)\end{array}\right)=E\left(\begin{array}{c}f_{j\uparrow}(r)\\
f_{j\downarrow}(r)\\g_{j\uparrow}(r)\\g_{j\downarrow}(r)\end{array}\right).\end{equation}
Next parameters appearing in this equation are
\begin{equation}F=\frac{sj}{1-\eta}-\frac{3}{2}\frac{\eta}{1-\eta}+\frac{1}{2},\hspace{1cm}C=\xi_x-\xi_y, \hspace{1cm}D=\xi_x+\xi_y.\end{equation}

\begin{figure}[htbp]
\includegraphics[width=160mm]{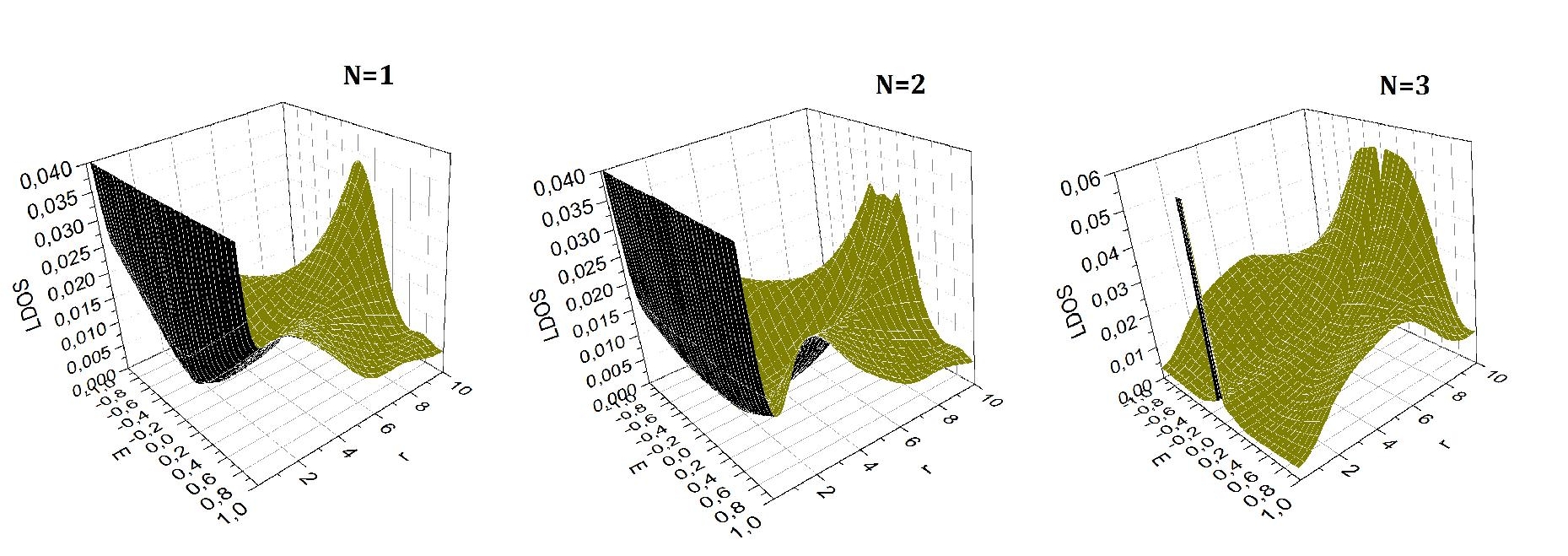}
\caption{3D graphs of $LDOS$ of the graphitic nanocone influenced
by SOC. Here, $LDOS$ corresponds to the sum of the solutions corresponding to $j=-1,0,1,2,3$. The number of the defects in the tip in the particular cases: $N=1$ (left), $N=2$ (middle) and $N=3$ (right).}\label{LDOS3Dspin}
\end{figure}

In \cite{soc}, a numerical method is introduced in detail which helps to find the solution of this system. Here, this method is outlined in Appendix. Using a modified version of Eq. (\ref{LDfin}) (we sum up the squares of absolute values of 4 components instead of 2 components), we calculate $LDOS$ from this solution. For different numbers of the defects in the conical tip, we see the resulting $LDOS$ in Figure \ref{LDOS3Dspin}. It involves the modes $j=-1,0,1,2,3$ with the same weight. While for the case of 1 and 2 defects in the tip, arbitrary energy and $r=0$, $LDOS$ grows to infinity, in the case of 3 defects in the tip, this effect appears close to zero energy only. Using a more thorough analysis, one could find out that the peak for the case of 3 defects corresponds to the case of the mode $j=-1$ and for other modes and the same number of defects, the behaviour is the same as in the case of 1 and 2 defects.

From these results follows that there could be a strong localization of the electrons in the tip, especially near zero energy in the case of 3 defects. Now, we will be interested, if this behaviour remains the same after the inclusion of some boundary effects which should simulate the real geometry of the nanostructure. Furthermore, we would like to ensure in this way the quadratical integrability of the solution.\\

\subsection{Incorporation of the boundary effects by a charge simulation}\

The influence of the charge considered in the conical tip is expressed in the Hamiltonian by the presence of the diagonal term $-\frac{\kappa}{r}$, where $\kappa=1/137$ is the fine structure constant. This term substitutes the diagonal zeros in Eq. (\ref{syst}):
\begin{equation}\label{syst2}\left(\begin{array}{cccc}-\frac{\kappa}{r} & 0 & \partial_r+\frac{F}{r} & -\frac{\rm i}{r}C\\0 & -\frac{\kappa}{r} & -\frac{\rm i}{r}D &
\partial_r+\frac{F}{r}\\ -\partial_r+\frac{F-1}{r} & \frac{\rm i}{r}D & -\frac{\kappa}{r} & 0\\
\frac{\rm i}{r}C & -\partial_r+\frac{F-1}{r} & 0 & -\frac{\kappa}{r}\end{array}\right)\left(\begin{array}{c}f_{j\uparrow,C}(r)\\
f_{j\downarrow,C}(r)\\g_{j\uparrow,C}(r)\\g_{j\downarrow,C}(r)\end{array}\right)=E\left(\begin{array}{c}f_{j\uparrow,C}(r)\\
f_{j\downarrow,C}(r)\\g_{j\uparrow,C}(r)\\g_{j\downarrow,C}(r)\end{array}\right).\end{equation}
In this way, the parallel influence of both the Coulomb interaction and SOC is considered \cite{boundaryNC}. To solve the resulting equation, we use the analogy of the numerical method used in \cite{soc} - this analogy is presented in \cite{boundaryNC}. From the calculated results, $LDOS$ is calculated using Eq. (\ref{LDfin}) again.

\begin{figure}[htbp]
\includegraphics[width=170mm]{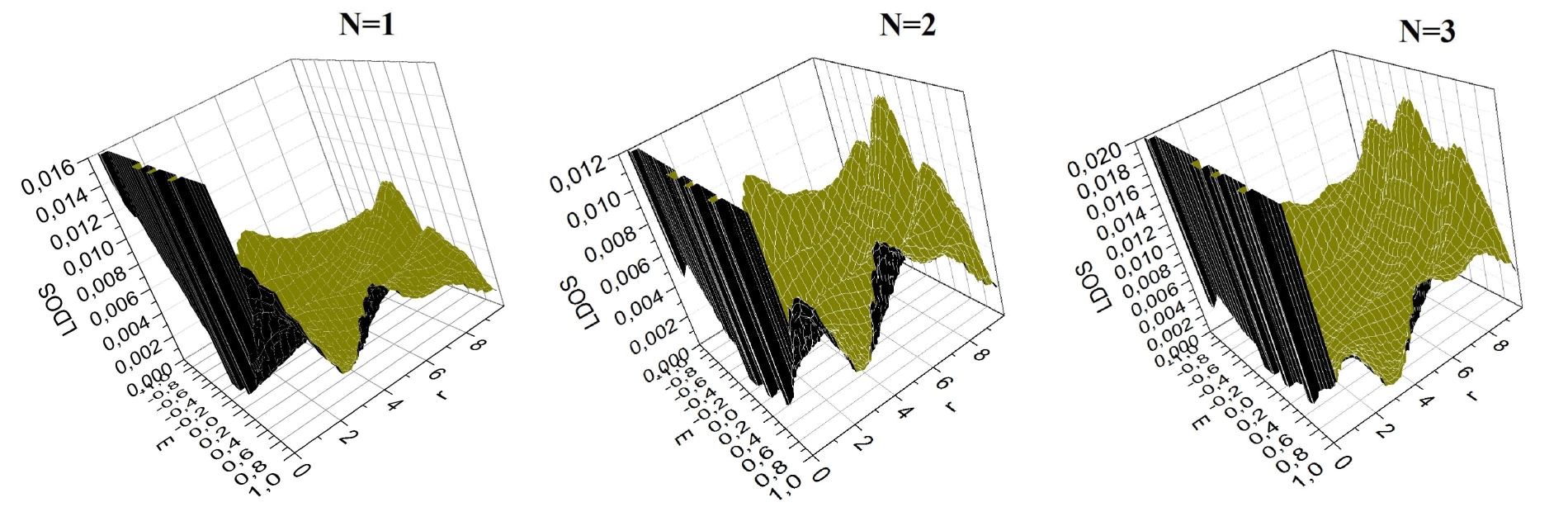}
\caption{Graphs of $LDOS$ of the graphitic nanocone influenced by the Coulomb interaction (including the influence of SOC) for different
distances $r$ from the tip, $-1\leq j\leq 3$ and for different numbers of the defects.}\label{graph-coulomb}
\end{figure}

The graphs of $LDOS$ based on the found solution are sketched in Figure \ref{graph-coulomb} for the same modes and numbers of the defects as in Figure \ref{LDOS3Dspin}, i.e. $-1\leq j\leq 3$. In spite of our expectations, this time the behaviour of the found result is the same for arbitrary number of the defects, i.e. the appearance and the uniqueness of the peak in the case of 3 defects is distorted.\\

\subsection{Comparison of the results}\

\begin{figure}[htbp]
\includegraphics[width=135mm]{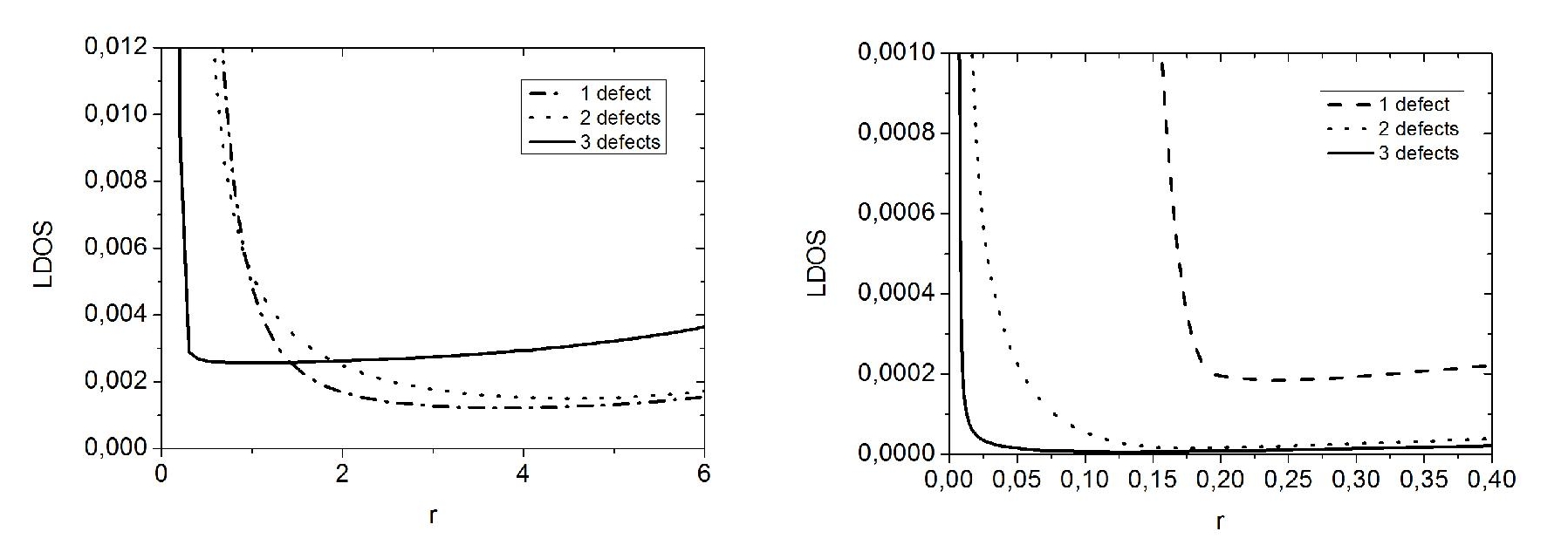}
\caption{Behaviour of $LDOS$ for zero energy close to $r=0$ for different numbers of defects in the conical tip: influence of SOC only (left) and the simultaneous influence of SOC and the Coulomb interaction (right).}\label{LDOSrCS}
\end{figure}

Now, we would like to verify the possible quadratic integrability of the solution found for the case of the additional effect coming from the charge simulation. In Figure \ref{LDOSrCS}, we see the dependence of $LDOS$ on $r$ variable close to zero energy for the case of the influence of SOC only and of the simultaneous influence of SOC and the Coulomb interaction. We see here that in comparison with the first case, in the second case the decrease of $LDOS$ close to $r=0$ is much faster and one could suppose that the quadratic integrability of the acquired solution is achieved here. To gain confidence with our conclusion, we have to do the integration of $LDOS$ in the investigated interval close to $r=0$ in all the outlined cases. This task is still in the progress.\\

\section{Properties of the graphitic wormhole}\

\begin{figure}[htbp]
\includegraphics[width=60mm]{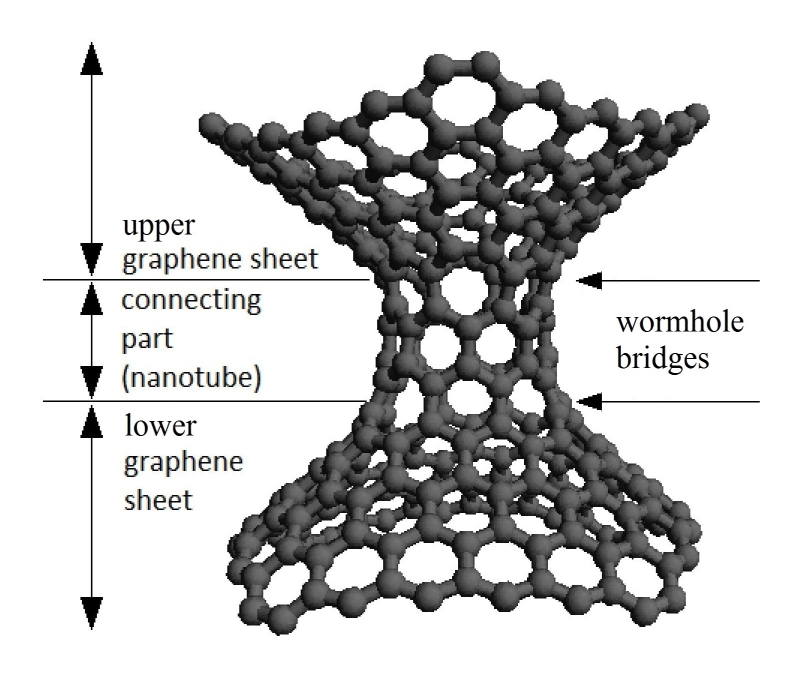}
\caption{The composition of the graphitic wormhole.}\label{compworm}
\end{figure}

The wormhole is understood as a form which arises when two graphene sheets are connected together with the help of the connecting nanotube. This can be achieved by the supply of the heptagonal defects onto both sides of the given nanotube. The number of the defects can vary from 1 to 12. The composition of the graphitic wormhole is depicted in Figure \ref{compworm}: it consists of the connecting nanotube and 2 (perturbed or unperturbed) graphene sheets. The places of the connections are called the wormhole bridges. Because of the physical limitations, the radius of the nanotube must be much larger than its length (this fact is ignored in Figure \ref{compworm} for the better illustration of the composition). The limit case of 12 defects is described in \cite{herrero1, herrero2}, in the other cases we speak about the so--called perturbed wormhole. Here, using the formalism of the subsection \ref{AS}, we derive the electronic structure for both cases and we will find out the form of the zero modes on the wormhole bridge. Furthermore, we investigate the influence of the additional effects which could appear here due to the extreme curvature in the place of the wormhole bridge - the relativistic mass acquisition of the present electrons. This effect, together with the effect of SOC which appears in the carbon nanotubes \cite{Ando} could lead to the appearance of the zero modes of the chiral massive electrons in the place of the wormhole bridge. This could serve as a useful instrument for the detection of the wormhole structures in the graphene bilayer during the process of the synthesis of the corresponding material.\\

\subsection{Electronic structure}\

We will solve Eq. (\ref{DirEq}) in the subsection \ref{AS}. In this case, the metric tensor has the form
\begin{equation}\label{metric}g_{\mu\nu}=\Lambda^2(r_{\pm})\left(\begin{array}{cc}1 & 0\\0 & r_{\pm}^2\end{array}\right),\hspace{5mm}
\Lambda(r_{\pm})=\left(a/r_{\pm}\right)^2\theta (a-r_{\pm})+\theta (r_{\pm}-a).
\end{equation}
Here, $\theta$ is the Heaviside step function, $r_-,r_+$ are the polar coordinates corresponding to the lower and the upper graphene sheet, respectively and $a=\sqrt{r_-r_+}$ is the radius of the wormhole.

\begin{figure}[htbp]
\includegraphics[width=75mm]{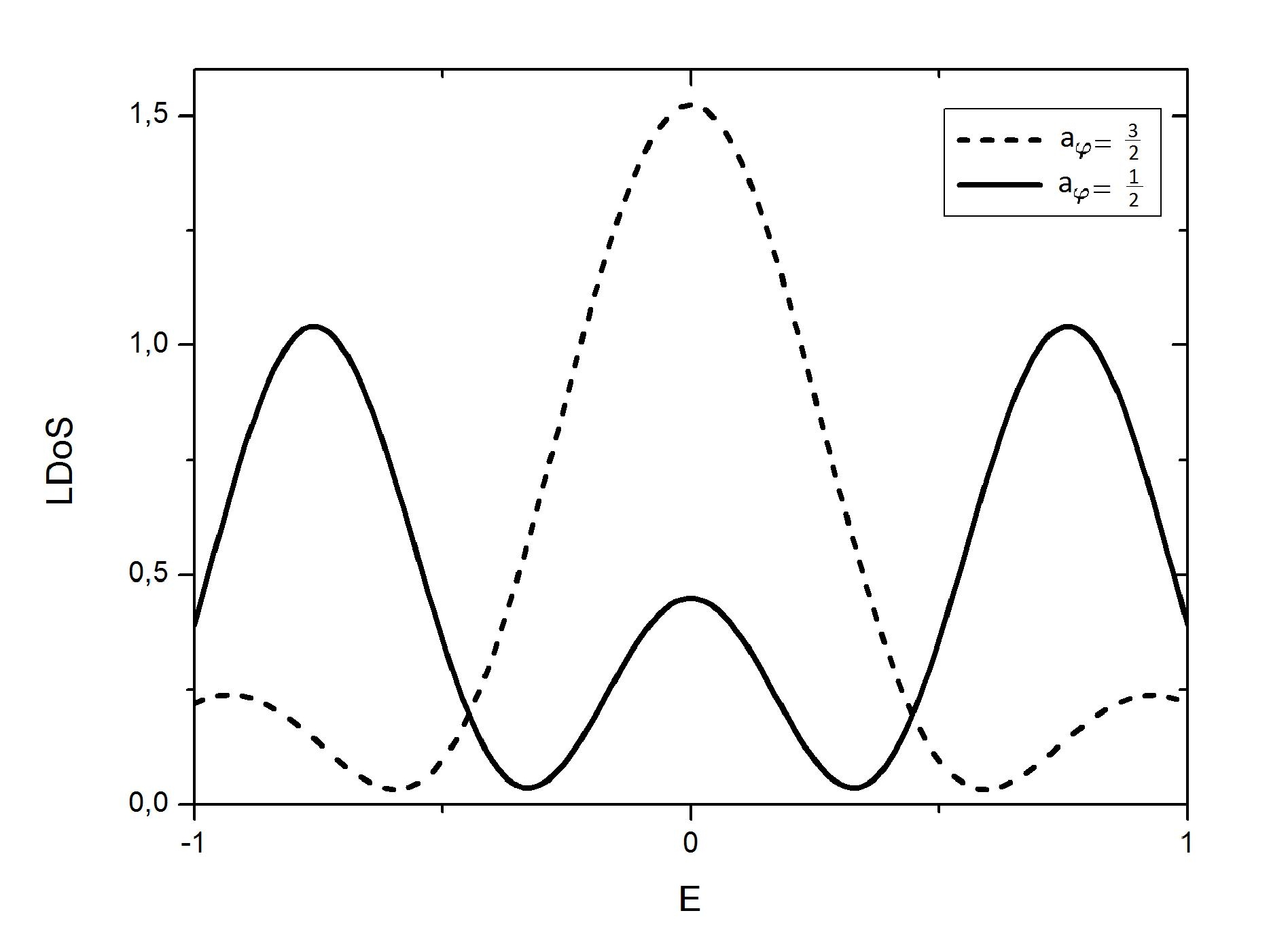}
\caption{Local density of states on the bridge of the graphitic wormhole for different values of $a_{\varphi}$.}\label{fg1}
\end{figure}

The values of the components of $a_{\mu}$ depend on the chiral vector \cite{dresselhaus} of the connecting nanotube. For our purpose, this vector is $(6n,6n)$ and $(6n,0)$. In most cases, $a_{\mu}$ has then the components
\begin{equation}a_{\varphi}=\frac{3}{2},\hspace{1cm}a_{r}=0.\end{equation}
The only exception is when the chiral vector is $(6n,0)$, where $n$ is not divisible by $3$. Then,
\begin{equation}a_{\varphi}=\frac{1}{2},\hspace{1cm}a_{r}=0.\end{equation}
Knowledge of the spin connection is also needed - the values of the components have the form
\begin{equation}\Omega_{\varphi}=-\frac{{\rm i}}{2}\sigma_3\left(r\frac{\Lambda'(r)}{\Lambda(r)}+1\right),\hspace{1cm}\Omega_r=0.\end{equation}
All these expressions we substitute into Eq. (\ref{DirEq}). The resulting equation is
\begin{equation}{\rm i}v_F\sigma^{\mu}(\partial_{\mu}+\Omega_{\mu}\mp{\rm i}\,a_{\mu})\psi^{\pm}=\varepsilon\psi^{\pm}.\end{equation}
Here, each sign $\pm$ corresponds to a different Dirac point (the corner of the reciprocal unit lattice). We get these 4 possibilities: for $r\geq a$,
\begin{equation}\label{geq}-{\rm i}v_F\left(\partial_r+\frac{1}{r}{\rm i}\partial_{\theta}\mp \frac{a_{\varphi}}{r}+\frac{1}{2r}\right)\psi_B^{\pm}=\varepsilon\psi_A^{\pm},\hspace{1cm} -{\rm i}v_F\left(\partial_r-\frac{1}{r}{\rm i}\partial_{\theta}\pm \frac{a_{\varphi}}{r}+\frac{1}{2r}\right)\psi_A^{\pm}=\varepsilon\psi_B^{\pm}\end{equation}
and for $0<r\leq a$,
\begin{equation}\label{leq}{\rm i}v_F\left(\frac{r}{a}\right)^2\left(\partial_r-\frac{1}{r}{\rm i}\partial_{\theta}\pm \frac{a_{\varphi}}{r}-\frac{1}{2r}\right)\psi_B^{\pm}=\varepsilon\psi_A^{\pm},\hspace{1cm} {\rm i}v_F\left(\frac{r}{a}\right)^2\left(\partial_r+\frac{1}{r}{\rm i}\partial_{\theta}\mp \frac{a_{\varphi}}{r}-\frac{1}{2r}\right)\psi_A^{\pm}=\varepsilon\psi_B^{\pm}.\end{equation}
In the first case, the solution is
\begin{equation}\psi^{\pm}=\left(\begin{array}{c}\psi^{\pm}_A(r,\varphi) \\ \psi^{\pm}_B(r,\varphi)\end{array}\right)=
c_1\left(\begin{array}{c}J_{j\mp a_{\varphi}-1/2}(kr) \\ -{\rm i\,sgn}\,\varepsilon J_{j\mp a_{\varphi}+1/2}(kr)\end{array}\right)+
c_2\left(\begin{array}{c}Y_{j\mp a_{\varphi}-1/2}(kr) \\ -{\rm i\,sgn}\,\varepsilon Y_{j\mp a_{\varphi}+1/2}(kr)\end{array}\right).
\end{equation}
Here, $J_j(x)$ and $Y_j(x)$ are the Bessel functions of the integer order $j$ and the energy $\varepsilon=\pm v_Fk$. To calculate $LDOS$, similarly as in the previous section, Eq. (\ref{LDfin}) is used. In Figure \ref{fg1}, different behaviour of $LDOS$, depending on the gauge field $a_{\varphi}$, is manifested.\\

\subsection{Zero modes}\

For the presented possibilities we investigate the zero modes - solutions of the Dirac equation for the zero energy. For this purpose, we consider zero values of the component $\psi_A^{\pm}$ of the solution. Then, from Eqs. (\ref{geq}), (\ref{leq}) follows: for $r\geq a$,
\begin{equation}\left(\partial_r-\frac{1}{r}{\rm i}\partial_{\theta}\mp \frac{a_{\varphi}}{r}+\frac{1}{2r}\right)\psi_B^{\pm}=0\end{equation}
and for $0<r\leq a$,
\begin{equation}\left(\partial_r-\frac{1}{r}{\rm i}\partial_{\theta}\pm \frac{a_{\varphi}}{r}-\frac{1}{2r}\right)\psi_B^{\pm}=0.\end{equation}
If $a_{\varphi}=\frac{3}{2}$ and $r\geq a$, the solution is
\begin{equation}\label{zero1}\psi_B^-(r,\varphi)\sim r^{-j-2}e^{{\rm i}\,j\varphi}.\end{equation}
The second possibility for this value of $a_{\varphi}$ is $0<r\leq a$, the corresponding solution is then
\begin{equation}\label{zero2}\psi_B^-(r,\varphi)\sim r^{-j+2}e^{{\rm i}\,j\varphi}.\end{equation}
Both solutions are strictly normalizable only for $j=0$. Analogous solution holds for $\psi_B^+$ and for $\psi_A^{\pm}$ if the components $\psi_B^{\pm}$ are chosen zero.

There are not strictly normalizable solutions for the value $a_{\varphi}=\frac{1}{2}$. It means, that in this case the zero modes do not exist.

\begin{figure}[htbp]
\includegraphics[width=120mm]{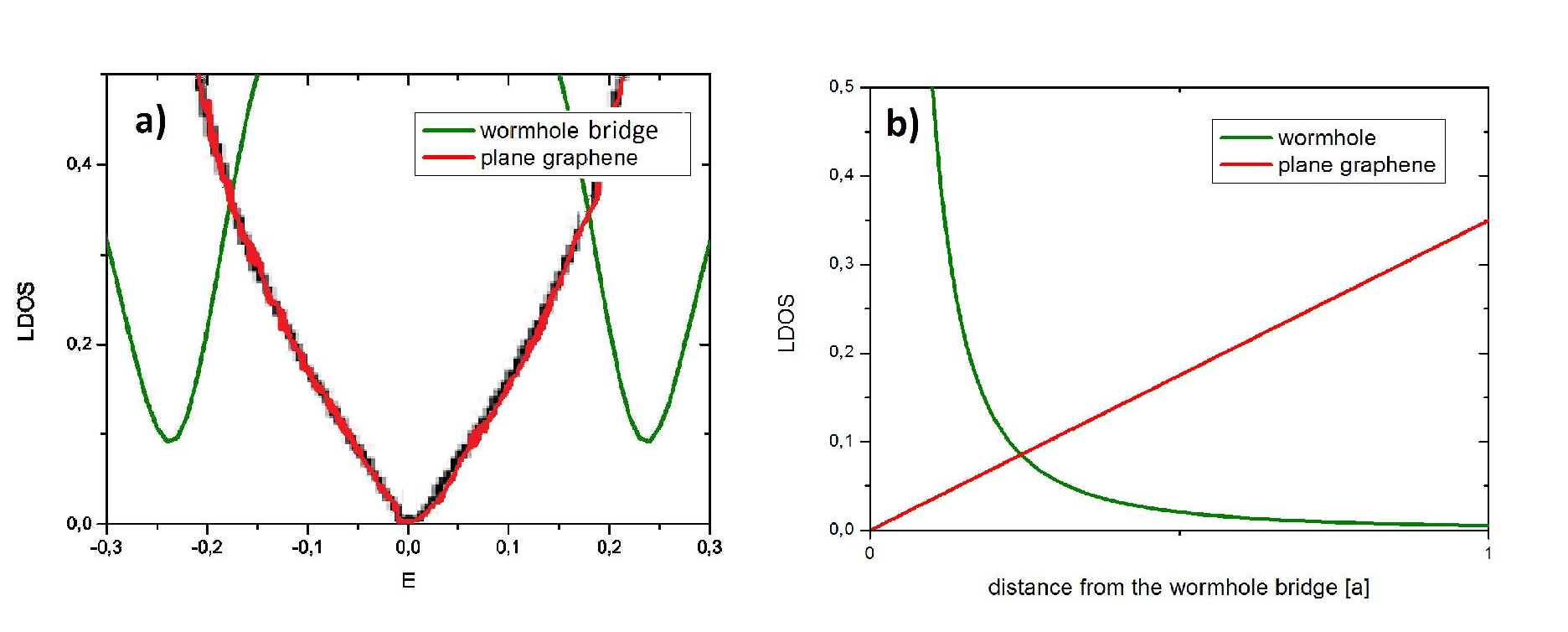}
\caption{Comparison of the properties of the wormhole and the plain graphene: a) local density of states, b) zero modes.}\label{fg2}
\end{figure}

On the base of these results, one could expect a strong localization of $LDOS$ near Fermi energy on the wormhole bridge. It is demonstrated in Figure \ref{fg2}a, where $LDOS$ of the plain graphene is supplied for the comparison. It could be experimentally observed. In Figure \ref{fg2}b, we see the comparison of the zero modes of these 2 structures at different distances from the wormhole bridge.\\

\subsection{Case of massive fermions}\

In the continuum gauge field-theory, zero mass of the fermions in the Dirac equation is considered (in other words, it is very small in comparison with energy). On the other hand, the extreme curvature of the investigated structure leads to such values of the Fermi velocity which cause the appearance of the relativistic effects. The changes of the Fermi velocity due to the curvature and other effects were demonstrated in \cite{kochetov,vf}. As the result, the mass of the fermions becomes considerable, similarly as in the bilayer graphene \cite{blvf0, blvf}. This effect is strengthened by the effective mass acquisition during the motion along the tube axis which happens due to the extreme size difference between the graphene sheets and the wormhole radius. This change of the space topology of graphene from 2D to 1D is similar to the string theory compactification. It means that we can image the wormhole connecting nanotube as the 1D object.

\begin{figure}[htbp]
\includegraphics[width=120mm]{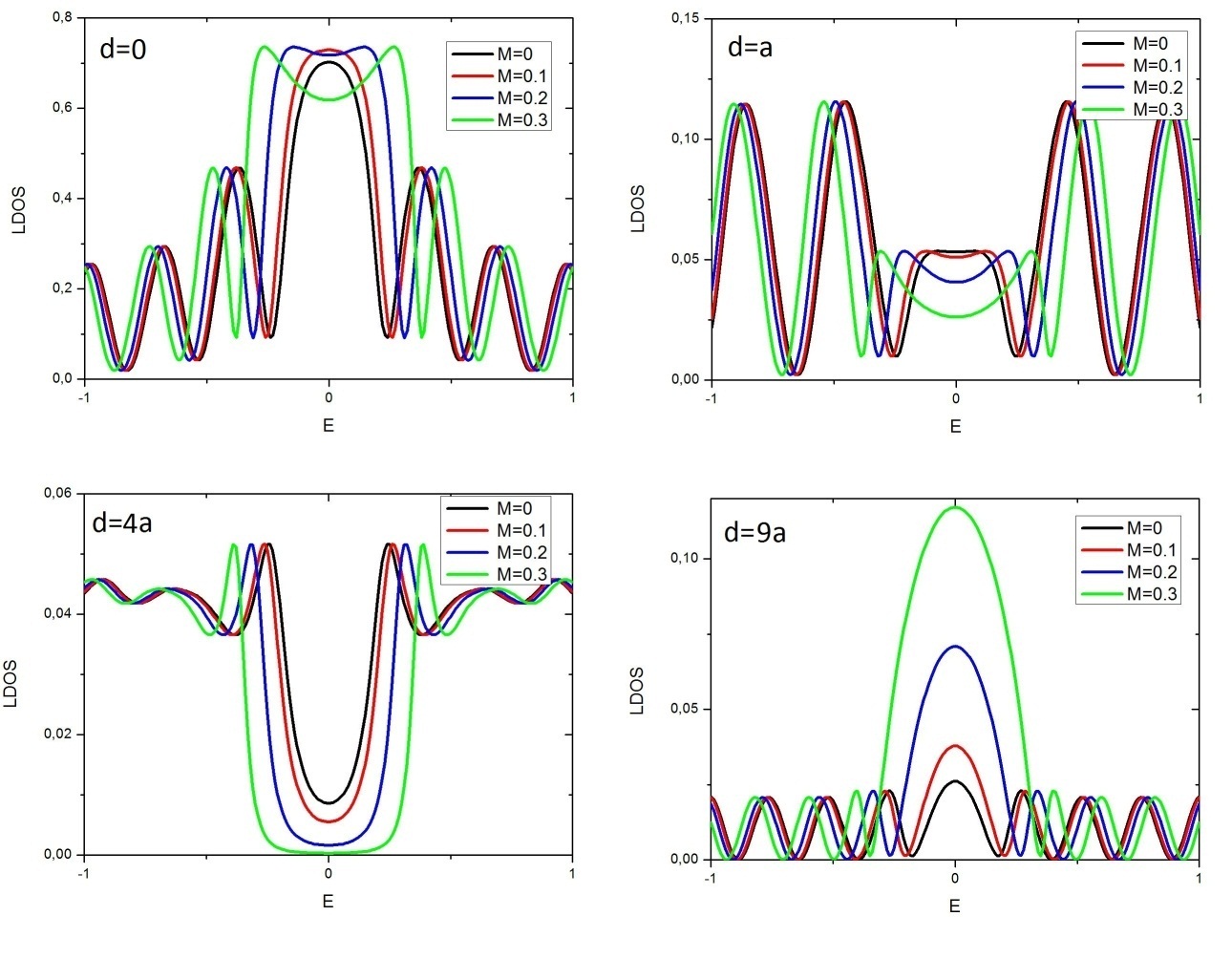}
\caption{Comparison of $LDOS$ for different masses of fermions at different distances $d$ from the wormhole bridge.}\label{fgMass}
\end{figure}

So, we need to incorporate a mass term into the Dirac equation (Eq. (\ref{DirEq})). To solve this problem, we go through the system of the corresponding equations (Eq. (\ref{Dsystem})) and transform it into the following differential equation of the second order:
\begin{equation}\label{secord}\left(\partial_{\xi\xi}-\frac{1}{2g_{\xi\xi}}\partial_{\xi}g_{\xi\xi}+\frac{\tilde{j}}{2}\sqrt{\frac{g_{\xi\xi}}
{g_{\varphi\varphi}^3}}\partial_{\xi}g_{\varphi\varphi}-\tilde{j}^2\frac{g_{\xi\xi}}{g_{\varphi\varphi}}+E^2g_{\xi\xi}\right)u_j=0.
\end{equation}
To simplify the calculations, the cylindrical geometry is supposed: the radius vector of the point at the surface changes as
\begin{equation}\vec{R}=(R\cos\varphi, R\sin\varphi, \xi),\end{equation}
where $R$ is the radius of the cylinder. In this case, Eq. (\ref{secord}) is considerably simplified:
\begin{equation}\left(\partial_{\xi\xi}+E^2-\frac{\tilde{j}^2}{R^2}\right)u_j=0,\end{equation}
which is solved by \cite{tubmass}
\begin{equation}u_j(\xi)=Ae^{k\xi}+Be^{-k\xi}.\end{equation}
Here,
\begin{equation}k=\sqrt{\frac{\tilde{j}^2}{R^2}-E^2}.\end{equation}
In \cite{ten, thal}, in a very similar form the dispersion relation is given for the massive 1D Dirac equation:
\begin{equation}k=\sqrt{M^2-E^2},\end{equation}
where $M$ is the mass of the corresponding fermion. From \cite{tubmass} an analogy indeed follows between the 2D massless and 1D massive case. On this base, we rewrite Eq. (\ref{secord}) into the form
\begin{equation}\label{massDir}\left(\partial_{\xi\xi}-\frac{1}{2g_{\xi\xi}}\partial_{\xi}g_{\xi\xi}+\frac{\tilde{j}}{2}\sqrt{\frac{g_{\xi\xi}}
{g_{\varphi\varphi}^3}}\partial_{\xi}g_{\varphi\varphi}-\tilde{j}^2\frac{g_{\xi\xi}}{g_{\varphi\varphi}}+(E^2-M^2)g_{\xi\xi}\right)u_j=0,
\end{equation}
in this case the mass $M$ corresponds to the fermion in the altered conditions. Now we find the corrections of $LDOS$ of the graphitic wormhole for different values of $M$. It is shown in Figure \ref{fgMass}. Our prediction is that these massive particles arising in the wormhole nanotubes could create energy bulks on the wormhole bridge and in the close area which should be experimentally measured by the STM or by the Raman spectroscopy \cite{enerbulk}. Moreover, this effect could be strengthened by the effect of SOC present in the connecting nanotube which was described in \cite{Ando} for the nanotubes and in section \ref{propNC} for the nanocone. This effect causes next energy splitting and as the result, the aforementioned chiral massive electrons could appear.

Another possibility to identify the wormhole structure comes from the fact that the massive particles could create strain solitons and topological defects on the bridge of the bilayer graphene which should propagate throughout the graphene sheet. These are almost macroscopic effects and should be caught by the experimentalists \cite{soliton}.\\

\subsection{Case of perturbed wormhole}\

Now we will investigate how the electronic structure changes if the number of the heptagonal defects on the wormhole bridge is lowered - in this way, the perturbed wormhole is created.

\begin{figure}[htbp]
\includegraphics[width=120mm]{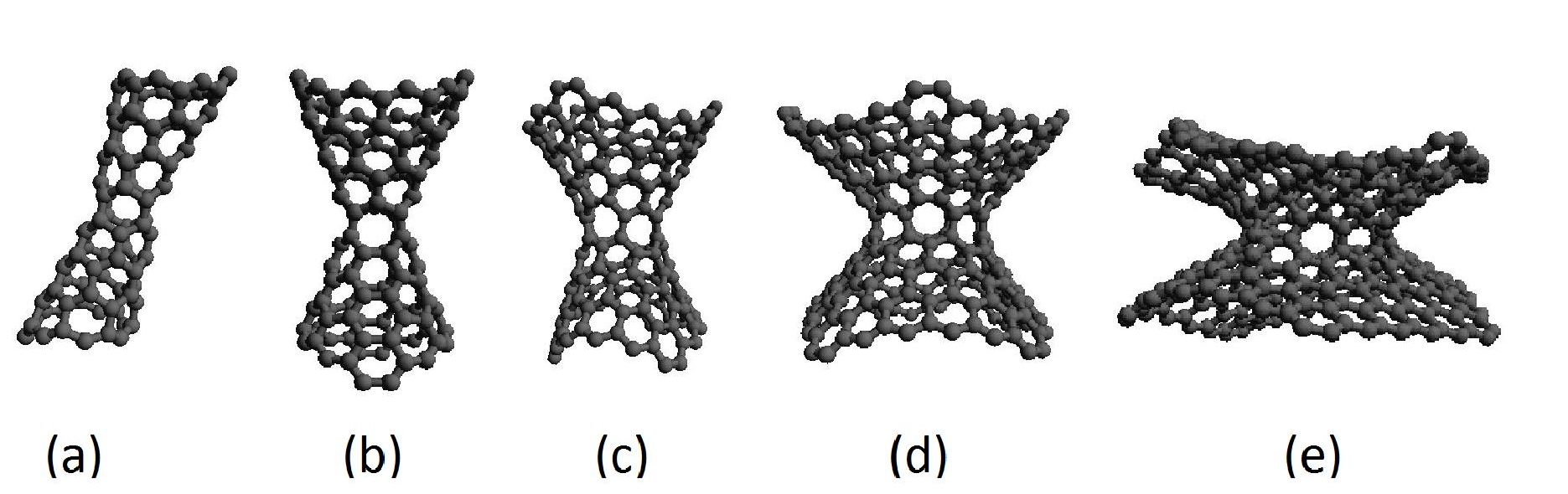}
\caption{Different forms of the perturbed wormhole: (a) 2 defects, (b) 4 defects, (c) 6 defects, (d) 8 defects, (e) 10 defects.}\label{fgwor}
\end{figure}

In Figure \ref{fgwor}, the possible forms of this structure are depicted. Due to symmetry preservation, only the even numbers of the defects, i.e. 2, 4, 6, 8 or 10, are considered.

\begin{figure}[htbp]
\begin{minipage}[b]{0.4\linewidth}
\centering
\includegraphics[width=60mm]{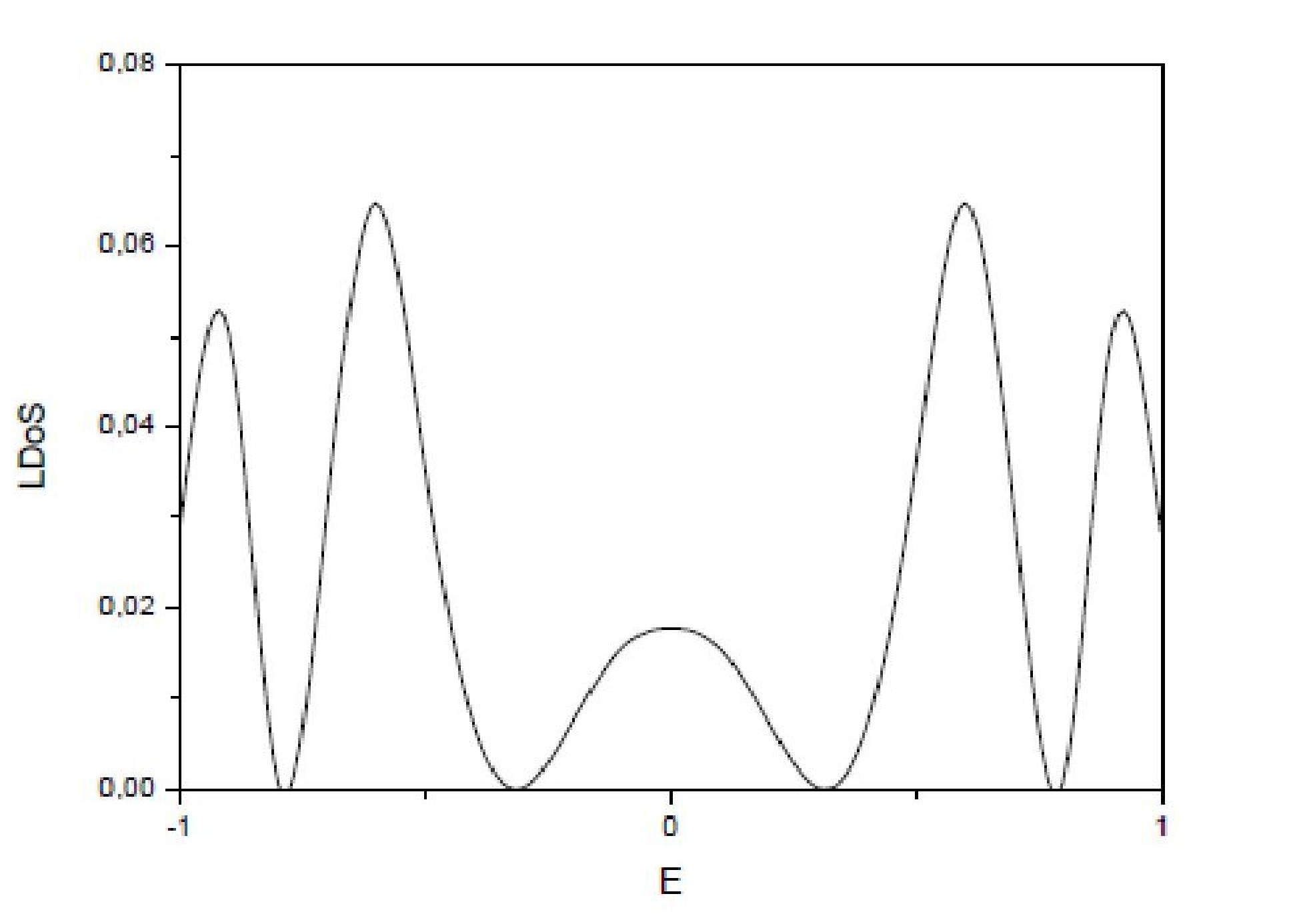}
\caption{Local density of states on the bridge of the graphitic perturbed wormhole.}\label{fg3}
\end{minipage}
\qquad\qquad
\begin{minipage}[b]{0.4\linewidth}
\centering
\includegraphics[width=50mm]{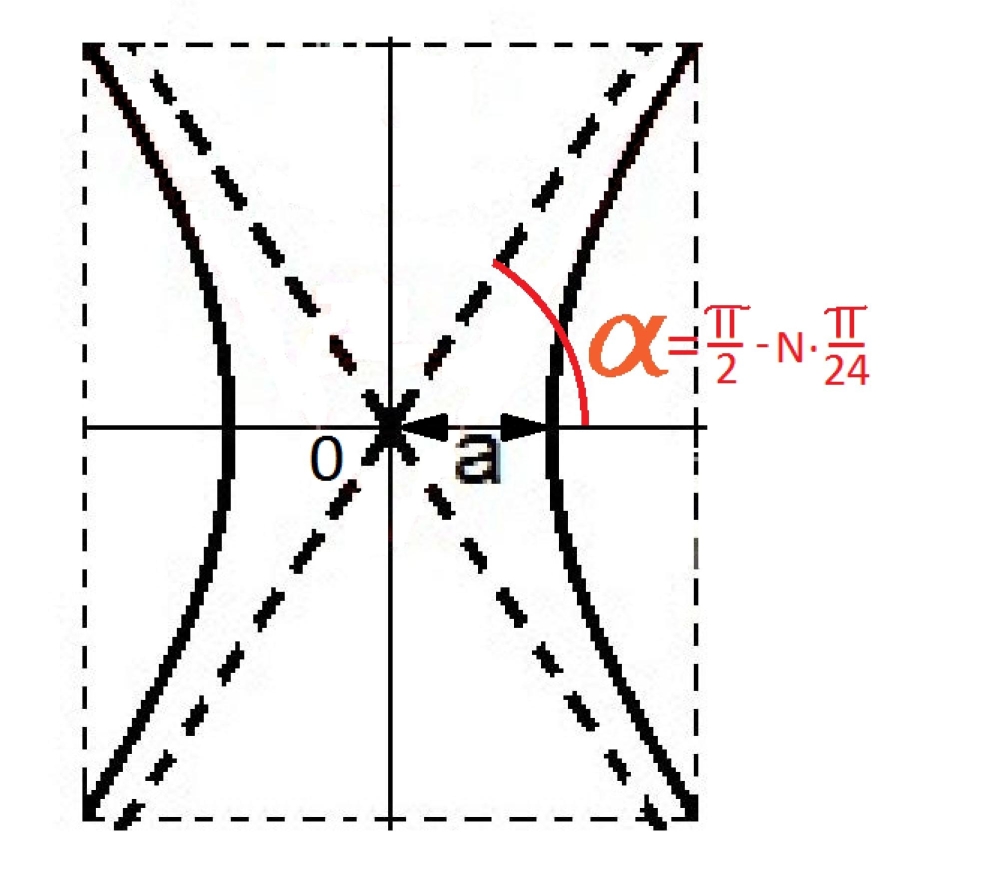}
\caption{Derivation of the $\triangle$ parameter.}\label{fgTriangle}
\end{minipage}
\end{figure}

The metric of the sheets can be draught by the radius vector
\begin{equation}\label{lvector}\overrightarrow{R}(z,\varphi)=\left(a\sqrt{1+\triangle z^2}\cos\varphi,
a\sqrt{1+\triangle z^2}\sin\varphi,z\right),
\end{equation}
where $\triangle$ is a positive real parameter; its value is derived from the number of the defects of the wormhole. In the case of $N=2$ defects, we can say that the value of this parameter is negligible, so $\triangle<<1$. Then, the nonzero components of the metric are
\begin{equation}g_{zz}=1+\frac{a^2\triangle^2 z^2}{1+\triangle z^2}\sim 1+a^2\triangle^2 z^2,\hspace{1cm}
g_{\varphi\varphi}=a^2(1+\triangle z^2).
\end{equation}
The nonzero components of the gauge fields are
\begin{equation}a_{\varphi}=N/4,\hspace{1cm}a_{\varphi}^W=-(2m+n)/3,\end{equation}
where $(n,m)$ is the chiral vector of the connecting nanostructure. Then, regarding the form of the spin connection and by the substitution into Eq. (\ref{DirEq}) we get the solution
\begin{equation}\psi_A(z)=C_{\triangle 1}D_{\nu_1}(\xi(z))e^{{\rm i}j\varphi}+C_{\triangle 2}D_{\nu_2}(i\xi(z)))e^{{\rm i}j\varphi},\end{equation}
\begin{equation}\psi_B(z)=\frac{C_{\triangle 1}}{E}\left(\partial_zD_{\nu_1}(\xi(z))-\frac{\widetilde{j}D_{\nu_1}(\xi(z))}{a}
(1-\frac{1}{2}\triangle^2z^2)\right)e^{-{\rm i}j\varphi}+
\frac{C_{\triangle 2}}{E}\left(\partial_zD_{\nu_2}(i\xi(z))-\frac{\widetilde{j}D_{\nu_2}(i\xi(z))}{a}
(1-\frac{1}{2}\triangle^2z^2)\right)e^{-{\rm i}j\varphi},\end{equation}
where
\begin{equation}\nu_1=i\frac{a^2\triangle-4 a^2 E^2+4ia\sqrt{\triangle}\widetilde{j}+4\widetilde{j}^2}{8a\sqrt{\triangle} \widetilde{j}},
\hspace{1cm}\nu_2=-i\frac{a^2\triangle-4 a^2 E^2-4ia\sqrt{\triangle}\widetilde{j}+4\widetilde{j}^2}{8a\sqrt{\triangle} \widetilde{j}},
\end{equation}
\begin{equation}\xi(z)= (-\triangle)^{1/4}\left(\sqrt{\frac{a}{
 2\widetilde{j}}} + \sqrt{\frac{2\widetilde{j}}{a}}z\right),
\end{equation}
$D_{\nu}(\xi)$ being the parabolic cylinder function. The functions $C_{\triangle 1}=
C_{\triangle 1}(E),\,C_{\triangle 2}=C_{\triangle 2}(E)$ serve as the normalization constants. We see the graph of the local density of states in Figure \ref{fg3}.

In the case of more than 2 defects, the value of $\triangle$ is non-negligible and we can get only the numerical approximation of $LDOS$. The derivation of the value of the parameter $\triangle$ follows from Figure \ref{fgTriangle}.

From this Figure follows that in the middle part, the upper branch of the graphene sheet converges to the line $z=x\cdot\tan\alpha$, where we can suppose that the angle $\alpha$ depends on the number of the defects $N$ linearly, i.e. $\alpha=\frac{\pi}{2}-N\cdot\frac{\pi}{24}$. (In this case, $\alpha=\frac{\pi}{2}$ corresponds to 0 defects and $\alpha=0$ corresponds to 12 defects.) Simultaneously, from Eq. (\ref{lvector}) follows that asymptotically we have
\begin{equation}\overrightarrow{R}(z\rightarrow\infty,\varphi)\rightarrow\left(a\sqrt{\triangle}z\cos\varphi,
a\sqrt{\triangle}z\sin\varphi,z\right),\end{equation}
from which follows
\begin{equation}z=x\cdot\tan\alpha=\left(a\sqrt{\triangle}\right)^{-1}x,\end{equation}
so
\begin{equation}\triangle=\frac{1}{a^2\tan^2\alpha}=\frac{1}{a^2\tan^2\left(\frac{\pi}{2}-N\cdot\frac{\pi}{24}\right)}.\end{equation}

\begin{figure}[htbp]
\includegraphics[width=120mm]{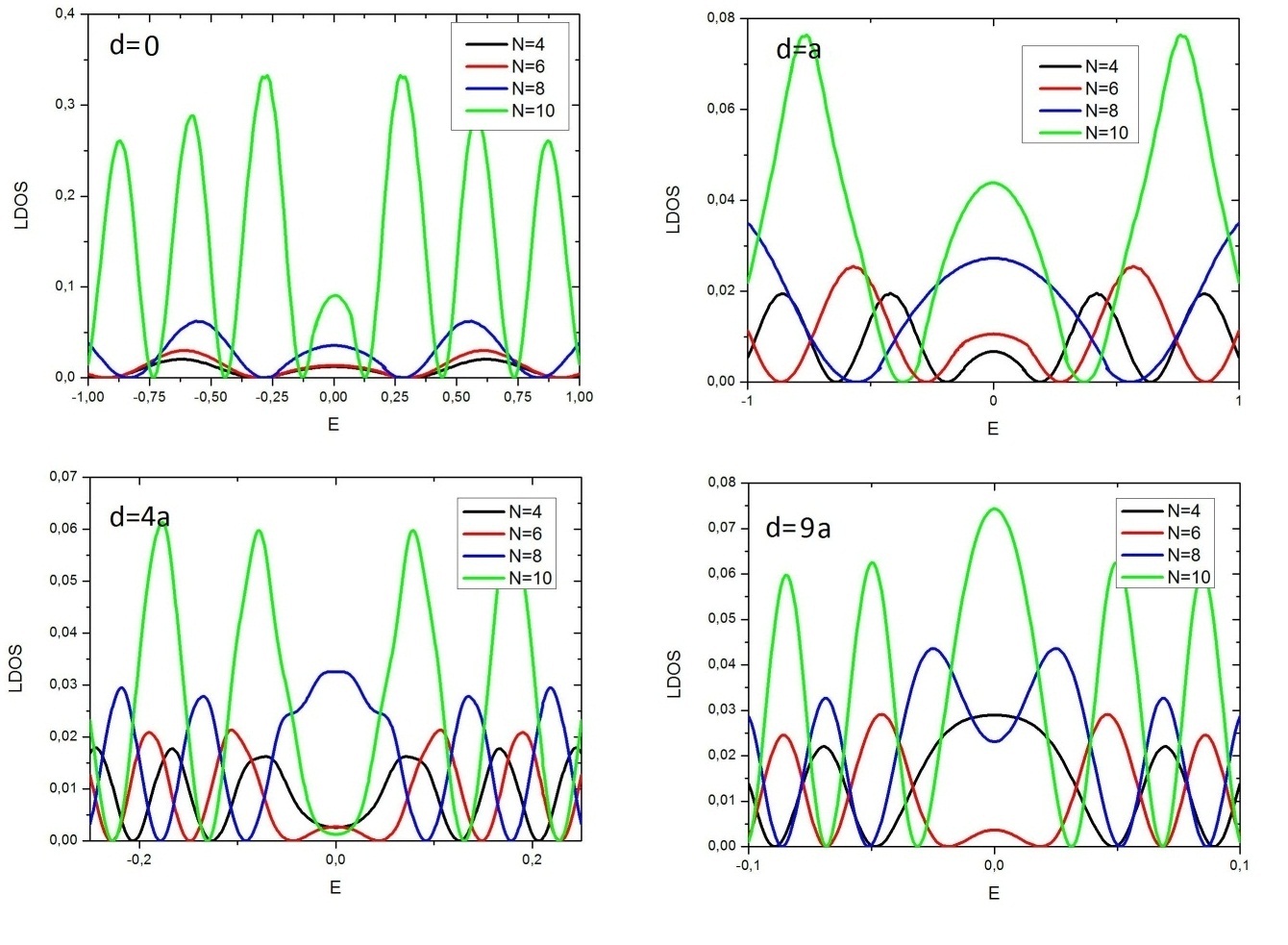}
\caption{Comparison of $LDOS$ for different numbers of the defects in the perturbed wormhole at different distances $d$ from the wormhole bridge.}\label{pertvar}
\end{figure}

In Figure \ref{pertvar}, we see the comparison of $LDOS$ for different kinds of the perturbed wormhole. From the plots follows that the intensity is rising with the increasing number of the defects and it is closer and closer approaching the results in Figure \ref{fg1}, where the case of 12 defects is shown.

\begin{figure}[htbp]
\includegraphics[width=80mm]{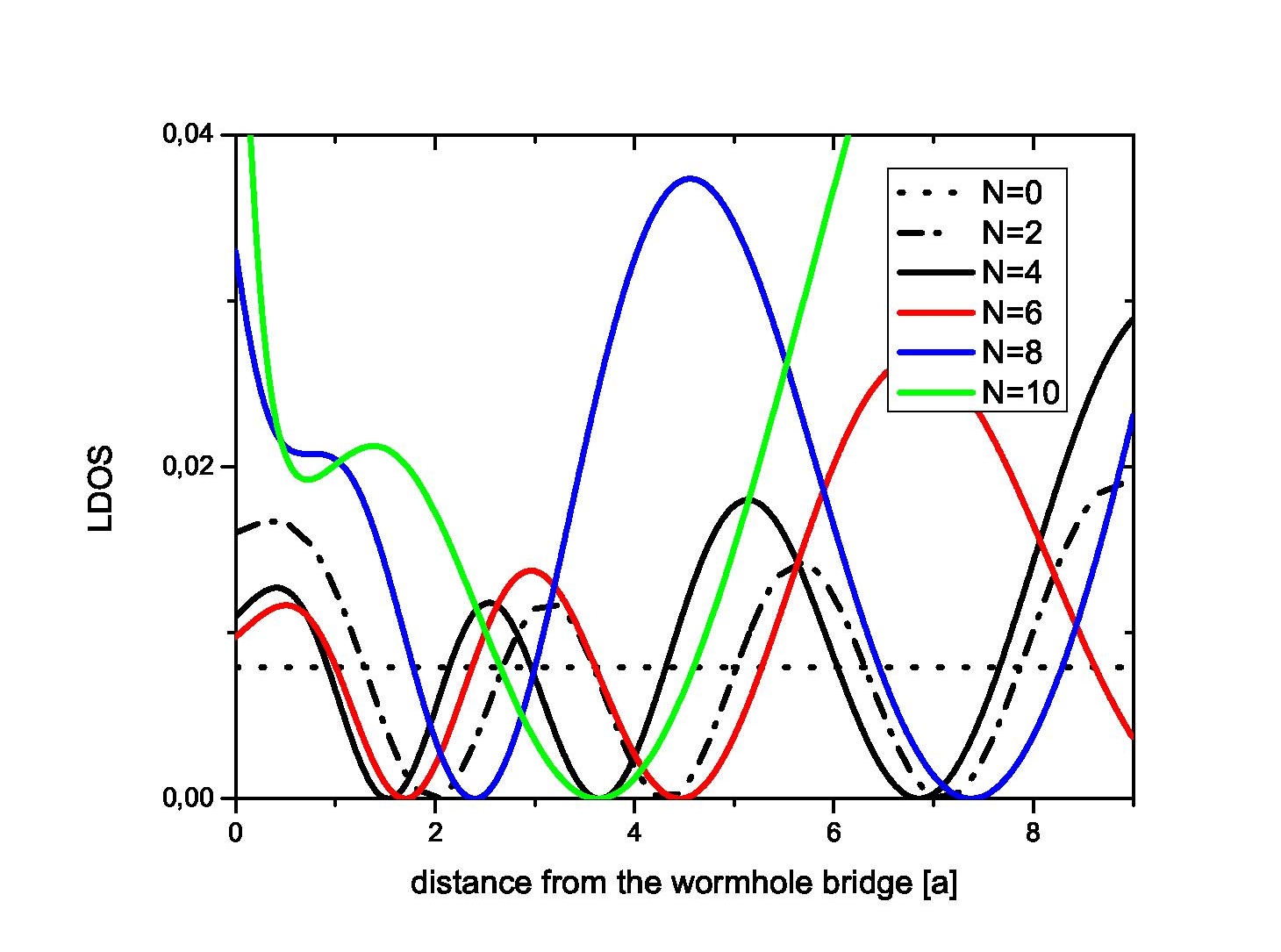}
\caption{Zero modes of the perturbed wormhole for different numbers of the defects.}\label{zeropvar}
\end{figure}

In Figure \ref{zeropvar}, $LDOS$ of zero modes is shown for a varying distance from the wormhole bridge in the units of the radius $a$ of the wormhole center. It was also acquired in the numerical way. For the unperturbed case (0 defects), the resulting plot resembles a line. In \cite{nanommta}, the exponential solution is found for this case but with a very slow increase, so, this could be that case. It is also seen from the plot that for the increasing number of the defects, the solution is approaching expressions in Eqs. (\ref{zero1}), (\ref{zero2}) for the zero modes of the unperturbed wormhole.

Of course, the massive fermions could also appear in the case of the perturbed wormhole. We will not perform a detailed derivation of the electronic structure for the case of this eventuality and we only note that the corrections to $LDOS$ would be an analogy of the corrections shown in Figure \ref{fgMass}.

\section{Conclusion}\

We performed the calculations of the electronic structure for the graphitic nanocone and the graphene wormhole. In the first case, our aim was to find the quadratically integrable solution which includes the boundary effects and considers the real geometry. This goal was partially achieved, but we need to verify the properties of the found solution close to the tip. The precision of the calculations could be improved by the better choice of the corresponding geometry, consideration of the discretion of the energetic spectrum coming from the finite size of the nanostructure and by the inclusion of next effects coming from the overlap of the neighboring atomic orbitals close to the tip \cite{nc2}. The localization of the electrons shown in Figures \ref{LDOS3Dspin} and \ref{graph-coulomb}, especially in the case of 3 defects, makes the graphitic nanocone a possible candidate for the construction of the scanning probe in atomic force microscopy.

In the second case of the graphene wormhole, we presented the mathematical motivation for our prediction of the effects which should appear close to the wormhole bridge. Our predictions will be verified with the help of the geometric optimizations and ab initio calculations. On this base, the most suitable candidates for the experiments will be chosen.

\appendix*

\section{Numerical Solution of the Dirac equation}\

To solve Eq. (\ref{syst}), we suppose the solution in the form of an infinite sum for each component:
\begin{equation}\label{solF}f_{j\uparrow}(r)=e^{\frac{\alpha}{r}+\beta r}\sum\limits_{k=0}^{\infty}a_kr^{\xi+k},\hspace{1cm}
f_{j\downarrow}(r)=e^{\frac{\alpha}{r}+\beta r}\sum\limits_{k=0}^{\infty}b_kr^{\xi+k},\end{equation}
\begin{equation}\label{solG}g_{j\uparrow}(r)=e^{\frac{\alpha}{r}+\beta r}\sum\limits_{k=0}^{\infty}c_kr^{\xi_1+k},\hspace{1cm}
g_{j\downarrow}(r)=e^{\frac{\alpha}{r}+\beta r}\sum\limits_{k=0}^{\infty}d_kr^{\xi_1+k}.\end{equation}
After the substitution into the corresponding system and comparison of the coefficients which correspond to the particular powers, we get $\xi=\xi_1-2$ and
\begin{equation}\label{eqA3}-\alpha c_0=Ea_0,\hspace{1cm}-\alpha c_1+\xi_1c_0+Fc_0-{\rm i}Cd_0=Ea_1,\hspace{1cm}-\alpha d_0=Eb_0,\hspace{1cm}-\alpha d_1+\xi_1d_0+Fd_0-{\rm i}Dc_0=Eb_1,\end{equation}
\begin{equation}\alpha a_0=0,\hspace{1cm}\alpha a_1+\xi a_0+(F-1)a_0+{\rm i}Db_0=0,\hspace{1cm}\alpha b_0=0,\hspace{1cm}\alpha b_1-\xi b_0+(F-1)b_0+{\rm i}Ca_0=0,\end{equation}
\begin{equation}\alpha a_2-\beta a_0+(F-\xi-2)a_1+{\rm i}Db_1=0,\hspace{1cm}\alpha b_2-\beta b_0+(F-\xi-2)b_1+{\rm i}Ca_1=0,\end{equation}
\begin{equation}\alpha a_3-\beta a_1+(F-\xi-3)a_2+{\rm i}Db_2=0,\hspace{1cm}\alpha b_3-\beta b_1+(F-\xi-3)b_2+{\rm i}Ca_2=0.\end{equation}
For the other indices, we get the system of the recurrence equations
\begin{equation}-\alpha c_k+\beta c_{k-2}+(F+\xi_1+k-1)c_{k-1}-{\rm i}Cd_{k-1}=Ea_k,\hspace{1cm}-\alpha d_k+\beta d_{k-2}+(F+\xi_1+k-1)d_{k-1}-{\rm i}Dc_{k-1}=Eb_k,\end{equation}
\begin{equation}-\alpha a_k+\beta a_{k-2}-(F-\xi_1+2-k)a_{k-1}-{\rm i}Db_{k-1}=-Ec_{k-4},\hspace{1cm}-\alpha b_k+\beta b_{k-2}-(F-\xi_1+2-k)b_{k-1}-{\rm i}Ca_{k-1}=-Ed_{k-4}.\end{equation}
If we suppose that $\alpha\neq0$, we get the zero solution. So, for the nontrivial solution $\alpha=0$ and as follows from the first and the third equation in Eq. (\ref{eqA3}), in this case the coefficients $a_0$ and $b_0$ must be also zero. Then, from the system
\begin{equation}(F-\xi_1)a_1+{\rm i}Db_1=0,\hspace{1cm}(F-\xi_1)b_1+{\rm i}Ca_1=0\end{equation}
follows:
\begin{equation}\xi_1=F\pm{\rm i}\sqrt{CD},\hspace{1cm}b_1=\pm\sqrt{\frac{C}{D}}a_1.\end{equation}
From the system
\begin{equation}(F+\xi_1)c_0-{\rm i}Cd_0=Ea_1,\hspace{1cm}(F+\xi_1)d_0-{\rm i}Dc_0=Eb_1,\end{equation}
we get
\begin{equation}c_0=\frac{(F+\xi_1)a_1+{\rm i}Cb_1}{(F+\xi_1)^2+CD}E,\hspace{1cm}d_0=\frac{Eb_1+{\rm i}Dc_0}{F+\xi_1}.\end{equation}
And from the system
\begin{equation}(F-\xi_1-1)a_2+{\rm i}Db_2=\beta a_1,\hspace{1cm}(F-\xi_1-1)b_2+{\rm i}Ca_2=\beta b_1,\end{equation}
follows
\begin{equation}b_2=\beta\frac{(F-\xi_1-1)b_1-{\rm i}Ca_1}{(F-\xi_1-1)^2+CD},\hspace{1cm}
a_2=\frac{\beta a_1-{\rm i}Db_2}{F-\xi_1-1}.\end{equation}
Here, $\beta$ is a free parameter. The following coefficients we get from the recurrence equations.\\

\end{document}